\documentclass[10pt,letterpaper]{IEEEtran}
\usepackage{amsmath}
\usepackage{amsthm}
\usepackage{amsfonts}
\usepackage{amssymb}
\usepackage{makeidx}
\usepackage{cite}
\usepackage{graphicx,bigints}
\usepackage{mathtools}
\usepackage{cases}
\usepackage{color}
\usepackage{hyperref}
\usepackage{bbm}
\usepackage[normalem]{ulem}
\usepackage{braket}
\usepackage{textcomp,gensymb}

\usepackage{enumitem}

\usepackage{epstopdf}
\usepackage{xcolor}
\usepackage{lipsum}

\usepackage{multirow}

\usepackage[makeroom]{cancel}
\makeatletter
\DeclareFontFamily{U}{tipa}{}
\DeclareFontShape{U}{tipa}{m}{n}{<->tipa10}{}
\newcommand{\arc@char}{{\usefont{U}{tipa}{m}{n}\symbol{62}}}%

\newcommand{\arc}[1]{\mathpalette\arc@arc{#1}}

\newcommand{\arc@arc}[2]{%
	\sbox0{$\m@th#1#2$}%
	\vbox{
		\hbox{\resizebox{\wd0}{\height}{\arc@char}}
		\nointerlineskip
		\box0
	}%
}
\makeatother

\newtheorem{theorem}{Theorem}

\newtheorem{example}{Example}

\newtheorem{proposition}{Proposition}
\newtheorem{remark}{Remark}

\DeclareMathOperator{\cO}{\mathcal{O}}

\DeclareMathOperator{\bS}{\mathbb{S}}

\DeclareMathOperator{\cR}{\mathcal{R}}

\DeclareMathOperator{\SNR}{\textnormal{SNR}}

\DeclareMathOperator{\bR}{\mathbb{R}}
\DeclareMathOperator{\bN}{\mathbb{N}}
\DeclareMathOperator{\bP}{\mathbf{P}}
\DeclareMathOperator{\ind}{\mathbbm{1}}
\DeclareMathOperator{\bE}{\mathbf{E}}

\newcommand*\diff{\mathop{}\!\mathrm{d}}

\newcommand*\nnb{\nonumber}

\newcommand{\overbar}[1]{\mkern 1.5mu\overline{\mkern-1.5mu#1\mkern-1.5mu}\mkern 1.5mu}

\newcommand{\ea}{\stackrel{(\text{a})}{=}}
\newcommand{\eb}{\stackrel{(\text{b})}{=}}

\definecolor{sandy}{HTML}{E6E2AF}
\definecolor{stone}{HTML}{A7A37E}
\definecolor{beach}{HTML}{EFECCA}
\definecolor{ocean}{HTML}{046380}
\definecolor{diver}{HTML}{002F2F}

\definecolor{awesome}{rgb}{1.0, 0.13, 0.32}

 \definecolor{azure}{rgb}{0.0, 0.5, 1.0}
 
 \definecolor{dollarbill}{rgb}{0.52, 0.73, 0.4}
 
\definecolor{Firenze1}{HTML}{468966}
\definecolor{Firenze2}{HTML}{FFF0A5}
\definecolor{Firenze3}{HTML}{FFB03B}
\definecolor{Firenze4}{HTML}{B64926}
\definecolor{Firenze5}{HTML}{8E2800}
\definecolor{mediumpersianblue}{rgb}{0.0, 0.4, 0.65}
\definecolor{hongik}{HTML}{004498}
\definecolor{cobalt}{rgb}{0.0, 0.14, 0.86}
\definecolor{burntorange}{rgb}{0.8, 0.33, 0.0}
\definecolor{hongik}{HTML}{004498}

\definecolor{guppiegreen}{rgb}{0.0, 1.0, 0.5}

\definecolor{ultramarineblue}{rgb}{0.25, 0.4, 0.96}

\title{Leveraging Aerial Platforms for Downlink Communications in Sparse Satellite Networks}

\author{Chang-Sik Choi~\IEEEmembership{Member~IEEE}
	\IEEEcompsocitemizethanks{\IEEEcompsocthanksitem{Chang-Sik Choi is an Assistant Professor of Dept. of EE, Hongik University, South Korea. email: chang-sik.choi@hongik.ac.kr}}
}
\begin{document}
	\hypersetup{pageanchor=false}
	
	\maketitle 
\begin{abstract}
Although a significant number satellites are deemed essential for facilitating diverse applications of satellite networks, aerial platforms are emerging as excellent alternatives for enabling reliable communications with fewer satellites. In scenarios with sparse satellite networks, aerial platforms participate in downlink communications, serving effectively as relays and providing comparable or even superior coverage compared to a large number of satellites. This paper explores the role of aerial platforms in assisting downlink communications, emphasizing their potential as an alternative to dense satellite networks. Firstly, we account for the space-time interconnected movement of satellites in orbits by establishing a stochastic geometry framework based on an isotropic satellite Cox point process. Using this model, we evaluate space-and-time performance metrics such as the number of orbits, the number of communicable satellites, and the connectivity probability, primarily assessing the geometric impact of aerial platforms. Subsequently, we analyze signal-to-noise ratio (SNR) coverage probability, end-to-end throughput, and association delay. Through examination of these performance metrics, we explicitly demonstrate how aerial platforms enhance downlink communications by improving various key network performance metrics that would have been achieved only by many satellites, thereby assessing their potential as an excellent alternative to dense satellite networks.
\end{abstract}
	
	
\section{Introduction}
\subsection{Related Work}
\IEEEPARstart{T}{hrough} circling the Earth, satellites in low Earth orbit (LEO) or medium Earth orbit (MEO) facilitate worldwide connectivity for terrestrial devices \cite{6934544,8700141}. Satellites find applications in various fields, including high-data-rate communications, integrated communication and sensing, large-scale data collection for Internet of Things, emergency low-power communications, and more\cite{8473417, 9526866, FCCKuiper,10144545}. Many challenges of satellite communications arise from the satellites rotating around the Earth at high speed. The speed of LEO satellites exceeds 8 km/sec, and with this high speed, the communication duration is limited to approximately 5 minutes for each satellite, as terrestrial gateways can only connect to the visible satellites. Because of this short communication duration, densely deploying satellites is deemed necessary to ensure continuous connection between satellites and terrestrial gateways \cite{8571192,9755278}.

When a large-scale dense deployment of satellites is infeasible for budgetary or practical reasons, an innovative alternative is employing aerial platforms in the air\cite{9502642,9711564,9861699,10239284}. Network operators may install high-altitude aerial platforms at distances closer to satellites to receive and transmit data with advanced equipment. These aerial platforms allow terrestrial gateways to communicate with satellites outside of their visibility scope, effectively extending the duration and range of downlink communications\cite{10239284}. In other words, the use of aerial platforms can be a highly effective alternative to dense satellite deployments, offering equivalent or better coverage with fewer satellites by overcoming the geometric limitations of sparse satellite networks.

Employing aerial platforms for downlink satellite communications affects network performance in both space and time domains. For instance, since aerial platforms are capable of covering larger areas in space, not only does the maximum communication distance increase, but the average communication duration also improves. The performance of aerial platform-assisted downlink satellite communication involves both space and time domain aspects, making it complex to comprehensively analyze under a unified framework equipped with spatial and temporal elements. To fairly evaluate the performance of such a network, it is necessary to build a systematic framework interconnecting the spatial and temporal components of the satellites, aerial platforms, and terrestrial gateways.

This paper seeks to uncover the potential of aerial platform-assisted satellite communications, particularly by establishing a space-time four-dimensional (4D) model based on a stochastic geometry framework \cite{daley2007introduction, baccelli2010stochastic, baccelli2010stochasticvol2, chiu2013stochastic}, and by analyzing key performance metrics closely linked to real-world challenges.

In the vast literature on the stochastic geometry models for satellite networks, binomial or Poisson models were employed by many papers to analyze space domain network performance \cite{9313025,9079921,9218989,9177073,9755277,9681887}. Specifically, these papers used the analytical models based on binomial or Poisson point processes to investigate performance metrics such as the typical interference and the signal to interference-plus-noise ratio (SINR) coverage probability. These papers produced expressions of network performance as functions of key network parameters, offering an efficient way of designing LEO satellite networks. However, neither model comprehensively incorporates the orbital foundation of satellite networks nor accounts for the clustering of satellites on orbits. The absence of orbital structure and satellite clustering may hinder a unified and accurate analysis of the space and time domain network performance of satellite networks assisted by aerial platforms.

Recently, \cite{choi2023cox} developed the Cox point process for LEO satellite networks, providing a novel stochastic geometric interpretation of their orbital structure. The presence of this orbital structure facilitates the analysis of both spatial and temporal aspects of LEO satellite networks. Specifically, \cite{choi2023cox} developed a Cox point process to describe satellites at various altitudes and derived its mathematical statistics, which are essential for analyzing LEO satellite networks. Building on these results, \cite{choi2023} determined the SINR coverage probability for downlink communications from satellites to ground users. They successfully demonstrated that the Cox point process locally approximates existing or forthcoming LEO satellite networks. More recently, \cite{choi2023hetero_COX} constructed an analytical framework to investigate open and closed access downlink communications based on heterogeneous satellite networks. Additionally, \cite{choi2023delay_COX} utilized the Cox model to explore a delay-tolerant data harvesting architecture leveraging LEO satellites. Although \cite{choi2023,choi2023hetero_COX} showed that the Cox model approximates any existing or future constellations by the moment-matching method, the Cox-based models do not consider any communication agents relaying messages from satellites to terrestrial gateways. Hence, the performance gains from aerial platforms predicted by \cite{9502642,9711564,9861699,10239284} cannot be mathematically analyzed, leaving the gains from aerial platforms not fully understood.

\begin{figure}
	\centering
	\includegraphics[width=.7\linewidth]{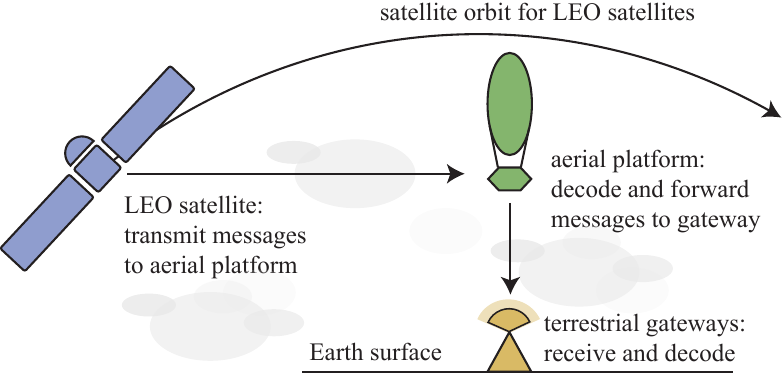}
	\caption{Aerial platform-assisted  satellite downlink communications.}
	\label{fig:networkarchitecture}
\end{figure}

Continuing the research on LEO satellite communications analysis, the present work pioneers the examination of aerial platform-assisted downlink communications from satellites to terrestrial gateways. This paper highlights the potential of such platforms by accurately modeling the dynamics of network elements and deriving the network performance in both the space and time domains. Ultimately, we aim to clearly demonstrate the benefits of introducing aerial platforms to sparse satellite networks, showcasing their role as an excellent alternative to densely populated satellite networks for downlink communications. Fig. \ref{fig:networkarchitecture} illustrates the network architecture we propose and analyze in this paper.

\subsection{Theoretical Contributions}
\subsubsection{Space-time Model for Satellite Dynamics and Downlink Communications} 
After deployment, LEO satellites form individual groups that travel along their designated orbits. To assess the performance of downlink communications from these satellites to aerial platforms, and eventually to terrestrial gateways, we employ a Cox-based stochastic geometry model to represent satellites' locations and motions. In this model, satellites are randomly distributed along orbital paths and maintain a steady velocity as they traverse these orbits. The terrestrial gateways, or equivalently terrestrial nodes, are randomly distributed on the Earth's surface. Above these terrestrial nodes, aerial platforms are positioned at a given altitude, playing a crucial role in facilitating downlink communications from satellites to terrestrial nodes. Specifically, they first decode messages from satellites and then forward the messages to the designated terrestrial nodes below them. The Cox satellite model employed in this paper distinguishes itself from conventional binomial or Poisson satellite models, where satellite points are not located on orbits; the absence of orbits in these models leads to an oversight of spatial correlation, resulting in less accurate analysis. By accurately modeling the locations and motions of satellites on their orbits, our approach facilitates the study of not only instantaneous network performance but also its variation over time, ensuring that the geometric impacts of aerial platforms are taken into account.

\subsubsection{Space and Time Performance Analysis} 
This paper investigates aerial platform-assisted satellite downlink communications from both space and time perspectives, utilizing the proposed model. Firstly, we accurately determine the average number of effective satellites from which a typical terrestrial gateway can directly receive messages. Subsequently, the paper evaluates the connectivity probability of the proposed network architecture, providing an accurate assessment of the increase in connectivity achieved with the help of aerial platforms. Under the proposed stochastic geometry framework, this derived metric also corresponds to the available time slots during which terrestrial nodes can receive downlink messages from satellites, directly and indirectly, out of the total time slots.

To further illustrate the benefits of aerial platforms, this paper derives the SNR coverage probability of the typical aerial platform and the typical terrestrial node, respectively, as functions of network parameters. Specifically, these formulas are presented as the complementary cumulative distribution function (CCDF) of the small-scale fading random variable, allowing for a swift assessment of various fading scenarios without the need for system-level simulations. By incorporating links from satellites to aerial platforms and links from aerial platforms to ground nodes, we define the end-to-end throughput of the proposed network architecture and illustrate its behavior by varying network variables. Finally, leveraging the proposed space-time unified structure, we calculate the minimum time required to establish a connection with any satellite and compare it to the scenario without aerial platforms, clearly demonstrating the role of aerial platforms in reducing network association time.

	\subsubsection{Key Design Insights to Real-World Problems}
This paper accurately derives key performance metrics, explicitly illustrating the geometric impact of aerial platforms. It is commonly believed that a large number of satellites is necessary to meet certain quality-of-service standards. However, by envisioning the use of aerial platforms to enable satellite downlink communications within a sparse satellite network, and by explicitly providing various network performance metrics both with and without aerial platforms, this work offers comprehensive guidance on using aerial platforms in a sparse satellite network with few satellites, thereby achieving comparable or even better network performance than a dense satellite network with many satellites. Finally, to better illustrate the applicability of the proposed architecture, we discuss operational and practical challenges, such as the elevation angle of aerial platforms and the impact of uneven terrain in the proposed aerial platform-assisted satellite network.


\section{System Model}
This section introduces the proposed system model for aerial platform-assisted  satellite downlink communications. We first present the spatial model, the operation and assistance of aerial platform, and finally the performance metrics of the proposed model.  

\begin{figure}
	\centering
	\includegraphics[width=0.6\linewidth]{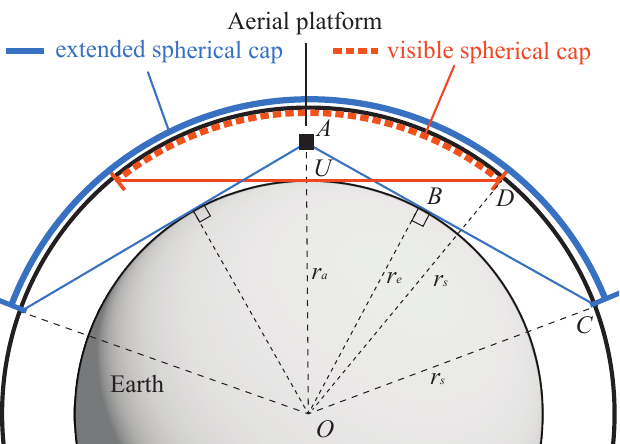}
	\caption{This figure visualizes the network connectivity improved by an aerial platform. The extended spherical cap (solid blue) represents the area visible from the aerial platform at $A$ whereas the visible spherical cap (dashed orange) represents the area visible from the typical terrestrial gateway at $U.$}
	\label{fig:fig1concept}
\end{figure}

\subsection{Satellites and Aerial Platforms Modeling}
We establish the Earth's center as the origin, with a radius denoted as $r_e=6371$ km. The $xy$-plane is the reference equatorial plane with the $x$-axis being the longitudinal zero point. The Earth's surface is $\bS_{r_e} = \{(x, y, z) \in \bR^3 \,|\, x^2 + y^2 + z^2 = r_e^2\}$ with $\bR^3$ being the Euclidean space.

\begin{figure}
	\centering
	\includegraphics[width=0.55\linewidth]{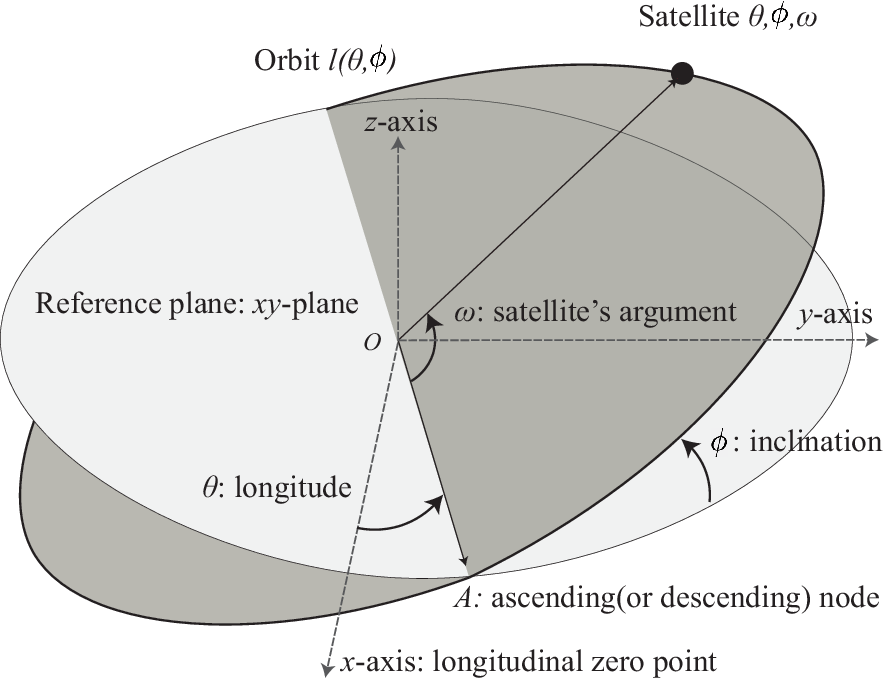}
	\caption{For the orbit (solid line), the longitude is $\theta,$ the inclination is $\phi$. Its satellite has an argument angle of $\omega$. The argument angle is the angle of the satellite from the point $A$, measured over the corresponding orbit.}
	\label{fig:fig2angles}
\end{figure}

To model the locations of satellites, we utilize a Cox point process that simultaneously generates both orbits and the satellites positioned upon them. \cite{choi2023cox}. Specifically, to generate orbits, we consider a Poisson point process $\Xi$ of density ${\lambda\sin(\phi)}/{(2\pi)}$ on a 2D rectangle set $\cR = [0, \pi) \times [0, \pi)$. The variable $\lambda$ corresponds to the mean number of orbits to be generated. Each point, denoted by $(\theta, \phi)$ of $\Xi$, corresponds to a single orbit $l(\theta, \phi) \in \bR^3$ of radius $r_s$, with longitude $\theta$ and inclination $\phi$. Refer to Fig. \ref{fig:fig2angles} for angles used in this paper. The Poisson orbit process on the rectangle is denoted by 
 \begin{equation}
 	\Xi = \sum_{i\in\bN} \delta_{\theta_i,\phi_{i}}. 
 \end{equation}
Equivalently, the orbit process on $\bR^3$ is denoted by 
 \begin{equation}
 	\cO = \bigcup_{(\theta_i,\phi_i)\in \Xi} l(\theta_{i},\phi_{i}).
 \end{equation}
Conditionally on the orbit process $\Xi$ (or equivalently $\cO$), the locations of LEO satellites are given by Poisson point processes $\{\psi\}$ of intensity $\mu/(2\pi r_s)$ on these orbits. The density parameter $\mu$ is the mean number of satellites per each orbit. Let $\psi_i$ be the Poisson point process restricted to the orbit $l(\theta_{i},\phi_{i})$. Then, collectively the satellite point process is denoted by 
\begin{equation}
	\Psi = \sum_{i\in\bN} \psi_i = \sum_{i\in\bN}  \sum_{j\in\bN} \delta_{X_{j,i}},
\end{equation}
where $X_{j,i}$ is the point representing the $j$-th satellite point on the orbit $l(\theta_i,\phi_i)$ where the satellite's argument angle is denoted by $\omega_j$. Since the satellite points are only on the orbit produced by the Poisson point process on the rectangle $\cR$, the proposed satellite point process is referred to as a Cox point process.

To account for the motion of satellites in space, we consider the fact that the satellites on each orbit move along the orbit on which they are located. Specifically, we assume that the angular speed of satellites is $\nu$ (rad/sec). At a given time $t$, the satellite with argument angle $\omega_j(t)$ on the orbit $l(\theta_i, \phi_i)$ has the $x, y, z$-coordinates as follows:
\begin{align*}
	x&= \sqrt{r_s^2\cos^2(\omega_j(t))+ r_s^2\sin^2(\omega_j(t))\cos^2(\phi_i)} \cos({\theta_i}+\tilde{\theta}),\\
	y&=\sqrt{r_s^2\cos^2(\omega_j(t))+ r_s^2\sin^2(\omega_j(t))\cos^2(\phi_i)} \sin({\theta_i}+\tilde{\theta}),\\
	z&=r_s\sin(\omega_j(t))\sin(\phi_i),\\
	\tilde{\theta}&=\arctan(\tan(\omega_j(t))\cos(\phi_i)).
\end{align*}
Here, $\Psi(t)$ denotes the satellite Cox point process at time $t$ and $\Psi$ denotes the satellite Cox point process at time zero.

\begin{figure}
	\centering
	\includegraphics[width=1\linewidth]{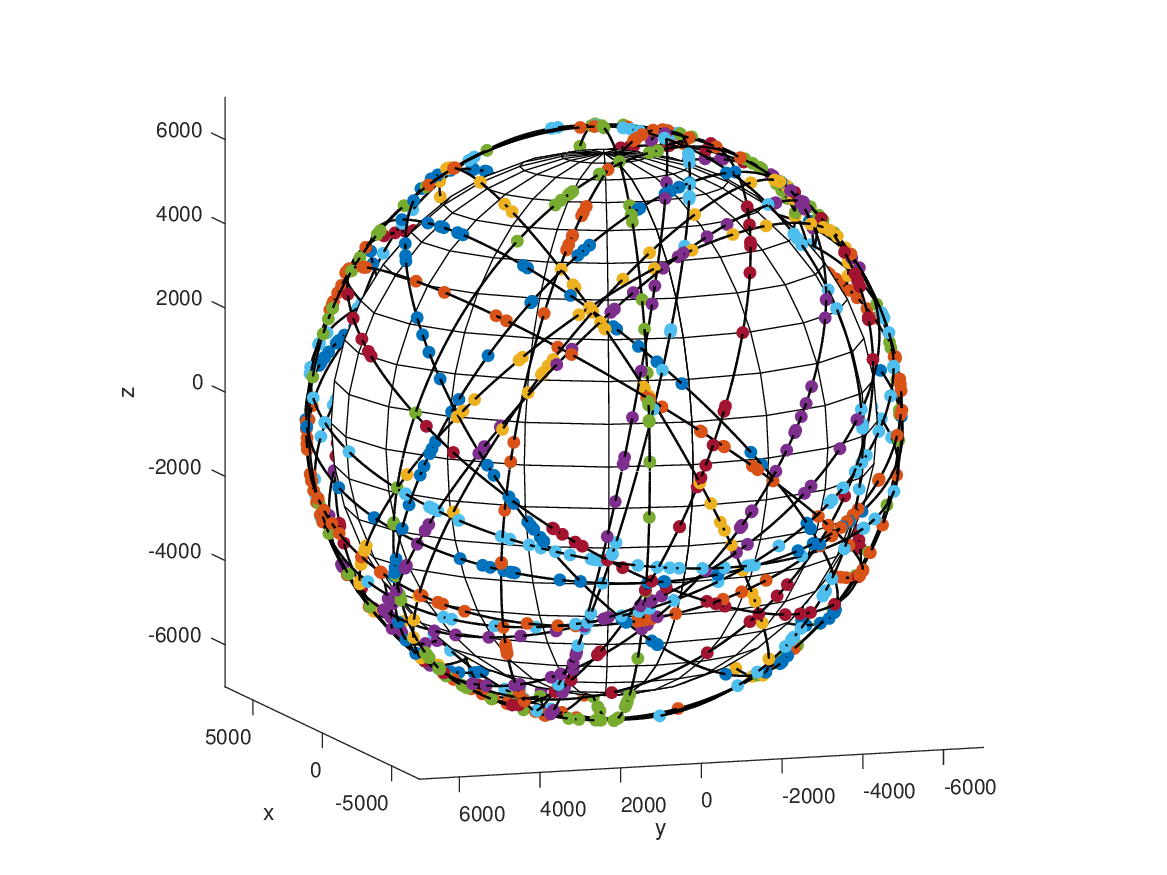}
	\caption{The proposed network with $\lambda=20$ and $\mu=50$.}
	\label{fig:fig3networklayout}
\end{figure}


It is crucial to acknowledge the geometric characteristics of the model: the points for satellites are only on the circles representing orbits. Fig. \ref{fig:fig3networklayout} effectively reproduces this geometric nature of the satellite networks. The orbits are illustrated by black solid lines, and the satellites are represented by colored dots. Note the proposed model is capable of generating a wide range of satellite layout scenarios. For instance, while maintaining the same mean number of satellites, one can jointly adjust the parameters $\lambda$ and $\mu$ to create a scenario with many-orbit less-satellite per orbit or a scenario with less-orbit many-satellites per orbit, and any other scenarios in between. 

To describe the locations of the terrestrial nodes for downlink communication receivers, we assume that $\lambda_g$ terrestrial nodes are randomly distributed on the Earth's surface. This paper focuses only on the terrestrial nodes assisted by aerial platforms. To maximize the applicability of the performance analysis presented in this paper, we make no restricting assumption such as the terrestrial nodes are uniformly distributed on Earth $\bS_{r_e}$.

Finally, aerial platforms are deployed to facilitate reliable communications from satellites to terrestrial nodes by relaying messages. We assume that the aerial platforms are located directly above the ground nodes, at the altitude of $h_a$ and that they first receive messages from satellites and then transmit those messages to the corresponding ground nodes. 

%

\subsection{Aerial Platform-Assisted Downlink Communications}
The aerial platforms are intermediate relaying nodes that enhance the reliability of downlink communications from satellites to ground nodes. We assume that every aerial platform receives downlink communication signals from its nearest satellite and then decodes the messages within. Subsequently, it forwards the message to its associated terrestrial gateway directly below the platform.\footnote{Various operations of aerial platforms can be considered within the proposed network geometry. To manifest the geometric impact of aerial platforms, we initially analyze the network performance while having aerial platforms as relays.}

We assume that every aerial platform is equipped with two phased antenna arrays with a gain of $g_a$. These two phased antenna arrays on each platform point directly to its associated satellite and its terrestrial gateway. Similarly, we assume phased antenna arrays on the satellites and ground nodes, each having gains of $g_s$ and $g_g$, respectively.

\subsection{Propagation Model}
For downlink communications, we assume a general model as in \cite{9681887,38821}. The received signal power of a receiver at distance $D\geq 1$ (meter) from the transmitter is $p g H D^{-\alpha}$ where $p$ is the received signal power at $1$ meter, $g$ is the aggregate antenna gain, $H$ is a random variable for small-scale fading, and $\alpha\geq 2$ is the path loss exponent. We initially consider a general Nakagami-m fading that create various fading scenarios by configuring its two parameters $m$ and $\Omega$. The probability density function (PDF) and the complementary cumulative distribution function (CCDF) of the random variable $H$ are provided as follows:
\begin{align}
	f_{{H}}(x;m, \Omega) &=\frac{x^{m-1}e^{-x/(m/\Omega)}}{\Gamma(m)(m/\Omega)^m}  \  \forall x\geq 0,\\
	P(H\geq x) &=1- \frac{\gamma(m, \Omega x/m)}{\Gamma(m)}   \ \forall x\geq 0,\label{fading}
\end{align}
respectively, where $\Gamma(\cdot)$ is the Gamma function and $\gamma(\cdot,\cdot)$ is the lower incomplete gamma function. 

Note that we will derive the formulas of the performance metrics as the general functions of the CCDF of the random variable $H$. In other words, our analysis easily incorporates any small-scale fading scenarios by simply replacing the above CCDF of $H$ with the CCDFs for the corresponding fading scenarios. See Example \ref{example1}.


\subsection{Performance Metrics}
We investigate the aerial platform-assisted downlink communications by evaluating the following network performance: effective spherical cap, connectivity probability, communication range distribution, SNR coverage probability, end-to-end network capacity, and delay distribution. In below, the performance metrics are briefly introduced. 

\par Throughout this paper, we position a typical gateway at the North pole $U=(0,0,r_e)$ and a typical aerial platform at $A=(0,0,r_a)$. Then we  analyze the network by taking the perspective of the typical gateway. Due to (i) the rotation invariant property inherent in the proposed satellite point process \cite{choi2023cox} and (ii) the spatial independence between the satellite point process and ground points, the network performance observed from the typical terrestrial gateway statistically represent the performance experienced by all ground nodes within the network. Likewise, the network performance observed from the standard aerial platform reflects the performance experienced by all aerial platforms within the network.

\begin{enumerate} 
\item Extended spherical cap: Without aerial platforms, the typical terrestrial gateway receives downlink signals only from its visible satellites. With the assistance of aerial platform, the typical terrestrial gateway can now get messages even from satellites outside its visible area. The typical extended spherical cap indicates the area where such satellites exist, or equivalently, the spherical cap visible from the typical aerial platform assisting the typical ground node. We analyze the average numbers of orbits and satellites in this extended spherical cap.

\item Connectivity: Through the aerial platforms, the typical terrestrial gateway can receive messages provided there is at least one satellite within its extended spherical cap. We examine network connectivity by deriving the probability that there is at least one satellite in the typical extended spherical cap at any given time. Within the proposed stochastic geometry framework, this connectivity probability corresponds to the fraction of time that a terrestrial gateway can get messages from any satellites.

\item Communication range: We assume that the typical aerial platform is associated with the nearest satellite. This paper derives the distance distribution from the typical aerial platform to its associated satellite, the nearest one in the extended spherical cap. The distance distribution is a key element determining the network performance since it gives the average amount of signal attenuation from satellites to aerial platforms.

\item SNR coverage probability: Since the aerial platforms decode the message from satellites and forward it to ground nodes, the reliability of the end-to-end communications of the proposed network will be enhanced. In this paper, we analyze the quality of the end-to-end communications by separately evaluating the SNR coverage probability of the typical aerial platform and the SNR coverage probability of the typical ground node.

		\item End-to-end throughput: Every aerial platform has two links to its associated terrestrial gateway and satellite, respectively. These two links jointly determine the throughput of the proposed network architecture. This paper defines the end-to-end throughput of the proposed network architecture as the minimum of the achievable rates of these two typical links connected by the typical aerial platform.

\item Geometric  association delay: When the number of satellites is very low, the typical terrestrial gateway may have no visible satellite and it may need to wait until a satellite appears. We define delay as the minimum time required for the typical terrestrial gateway  to wait until a satellite appears within the spherical cap, as explained above. We derive the association delay with and without aerial platforms, respectively, as functions of key parameters and compare them to demonstrate the benefits of using aerial platforms.
\end{enumerate}

\subsection{Preliminary}\label{S:prelim}
\subsubsection{Time invariant} The proposed network model is time-invariant, namely $\Psi(t) \stackrel{d}{=}\Psi $ since (i) the satellites conditionally on orbits are time-invariant Poisson point processes \cite{baccelli2010stochastic}  and (ii) the orbit process is rotation invariant \cite{choi2023cox} and the time passage can be represented as the rotation of orbits and their corresponding satellite points. For detail, see \cite{choi2023cox,choi2023,choi2023hetero_COX,choi2023delay_COX}. 
\subsubsection{Spherical cap} For the evaluation of the network performance, this paper uses a spherical cap $\bS_{X,d}$  the collection of points on a sphere whose Euclidean distances from a certain location $X$ are less than a given value $d$. The spherical cap with two variables $X$ and $d$ is defined as follows: 
\begin{equation}
	\bS_{X,d} = \{(x,y,z)\in\bS_{r_s} | \|(x,y,z)-X\|\leq d\},
\end{equation}
where $X$ is the target point and it is either $U$ the location of the typical terrestrial gateway or $A$ the location of the typical aerial platform. From geometry, $d \in [r_s-\|X\|, r_s+\|X\|]$. For instance, ${\bS}_{U,r_s-r_e}$ is the subset of the sphere of radius $r_s$ such that the distances from $U$ are less than $r_s-r_e$ and in this case, it is a single point $(0,0,r_s)$. Similarly, ${\bS}_{U,r_s+r_e} $ is the subset of the sphere $\bS_{r_s} $ such that the distances from $U$ is less than $r_s+r_e $ and in this case, it is $\bS_{r_s}$.

\subsubsection{Arc length} For the spherical cap ${\bS}_{A,d}$, the orbit $ l(\theta,\phi)$ intersects the spherical cap if $\phi\in[\pi/2-\zeta(d),\pi/2+\zeta(d)]$ where the critical angle $\zeta(d)$ is derived in \cite{choi2023cox} as follows: 
\begin{equation}
	\zeta({d}) = \arccos((|\overbar{OA}|^2+r_s^2-d^2)/(2r_s\|\overbar{OA}\|)).
\end{equation}

When the orbit $l(\theta,\phi)$ meet the cap ${\bS}_{A,d}$, the arc length for the intersection of the orbit $l(\theta,\phi)$ and the cap ${\bS}_{A,d}$ is 
\begin{equation}
	2r_s\arcsin(\sqrt{1-\cos^2(\zeta(d))\csc^2(\phi)}).
\end{equation}
The central angle corresponding to this arc is given by  $2\arccos(\sqrt{1-\cos^2(\zeta(d))\csc^2(\phi)}).$

\subsubsection{Distance to sphere}Let $X_{j,i}$ be the $j$-th point of the $i$-th orbit where the satellite $X_{j,i}$'s argument angle is $\omega_j$ and the orbit's inclination angle is $\phi_i.$ Then, a simple geometric argument shows that the distance from $Z$ on the positive $z$-axis to the satellite $X_{j,i}$, namely $|\overbar{ZX_{j,i}}|$ is given by 
\begin{equation}
	|\overbar{ZX_{j,i}}| =\sqrt{r_s^2-2r_s\|Z\|\sin(\omega_j)\sin(\phi_i) + \|Z\|^2}. \label{Eq.dist}
\end{equation}

\begin{table}
	\centering
	\caption{Notation}
	\begin{tabular}{|c|c|}
		\hline
		Variable & Description \\
		\hline 
		$\bS_{r_e}$ & Earth surface $\{(x,y,z)|x^2+y^2+z^2=r_e^2\}$\\
		\hline 
		$\bS_{r_s}$ & Satellite surface  $\{(x,y,z)|x^2+y^2+z^2=r_e^2\}$ \\
		\hline
		$\Xi$ &Poisson point process on $\cR$  \\
		\hline
		$l(\theta,\phi)$& Orbit  of radius $r_s$ with longitude $\theta$ and inclination $\phi$\\
		\hline
		$\cO$  & Isotropic orbit process on $\bR^3$  \\
		\hline
		$\psi_i$& Satellite point process on the orbit $l(\theta_i,\phi_i)$  \\
		\hline 
		$X_{j,i}$ &$j$-th satellite with argument $\omega_j$ on $i$-th orbit \\
		\hline
		$U=(0,0,r_e)$ & Typical user location \\
		\hline 
		$A=(0,0,r_a)$ & Typical aerial platform location\\
		\hline 
		$\bS_{X,d}$ & Spherical cap with distance less than $d$ from $X$	\\
		\hline 
	\end{tabular}
\end{table}

\section{Performance Analysis}\label{Section:result}
\subsection{Effective Orbits and Satellites}
With the assistance of aerial platforms, the ground nodes are able to get message from both visible and invisible satellites. This section analyzes the typical extended spherical cap by deriving the average numbers of orbits and satellites therein. Such orbits and satellites are referred to effective orbits and effective satellites, respectively, since the typical terrestrial gateway can now get messages from such satellites on those orbits. 

\begin{figure}
	\centering
	\includegraphics[width=1\linewidth]{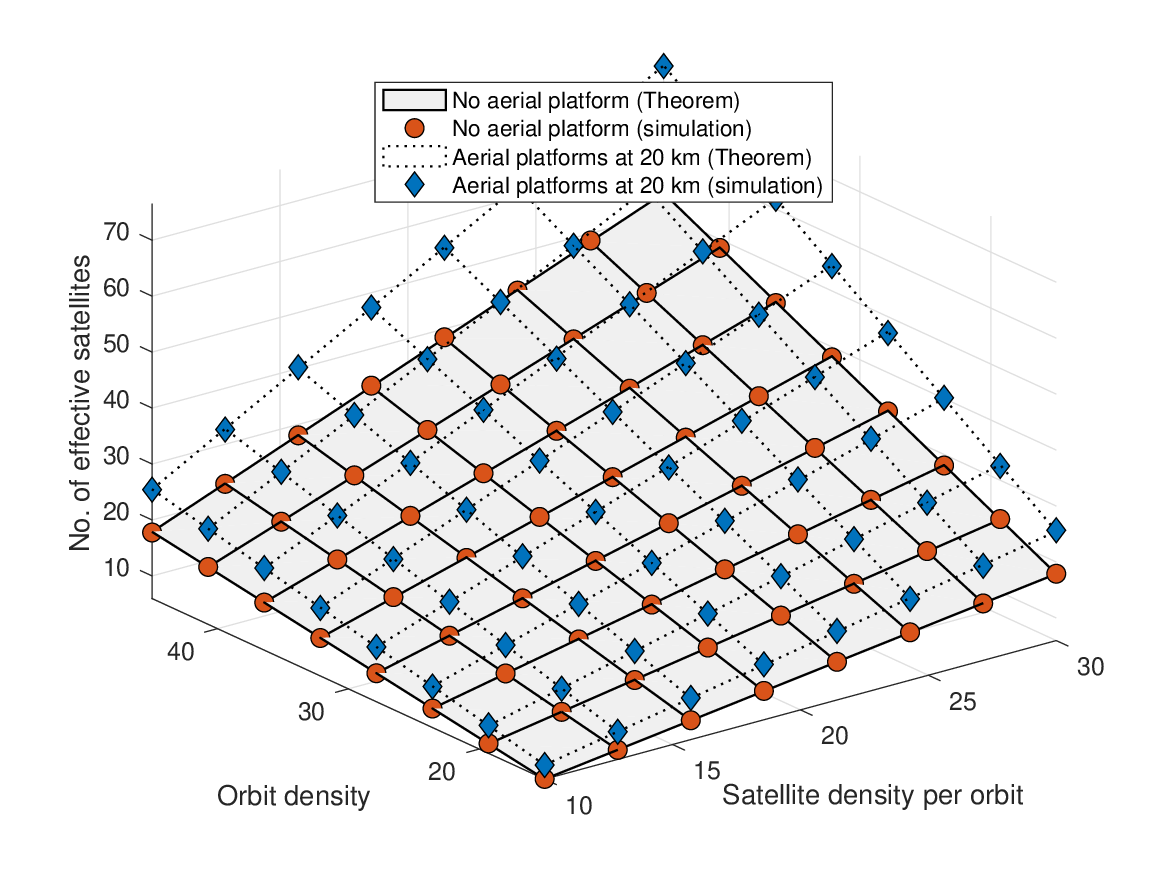}
	\caption{The circle marks indicate the average number of effective satellites without aerial platforms whereas the diamond marks indicate the average number of effective satellites with aerial platforms at the altitude of $20$ km.}
	\label{fig:theorem1}
\end{figure}

\begin{theorem}\label{Theorem1}
In the typical extended spherical cap, the average number of effective orbits is $ \lambda\sin(\overbar{\varphi}) $ and the average number of effective satellites is given by 
\begin{equation}
	\frac{\lambda\mu }{\pi}\int_0^{\overbar{\varphi}} \cos(\varphi){\arcsin(\sqrt{1-\cos^2(\overbar{\varphi})\sec^2(\varphi)})}\diff \varphi,
\end{equation}
respectively, where $\overbar{\varphi} = \arccos(r_e/r_a) + \arccos(r_e/r_s) $ and $r_e<r_a<r_s.$
\end{theorem}
\begin{IEEEproof}
	See Appendix \ref{A:1}
\end{IEEEproof}
In below, we evaluate the numbers of effective orbits and satellites without aerial platforms. This shows the benefits of employing aerial platforms. 
\begin{example}
	Without aerial platforms,  orbits are visible only if they intersect the spherical cap ${\bS}_{U,\overbar{UD}}$ (Fig. \ref{fig:fig1concept}). The inclinations of such orbits are in the interval $ \phi\in(\pi/2-\overbar{\xi},\pi/2+\overbar{\xi})$ where $\overbar{\xi} = \arccos(r_e/r_s).$ In this case, the average number of effective orbits is given by 
	\begin{align}
		 \int_{0}^{\pi}\int_{\pi/2-\overbar{\xi}}^{\pi/2+\overbar{\xi}}\frac{\lambda\sin(\phi)}{2\pi}\diff \theta\diff \phi= \lambda\sin(\overbar{\xi}). 
	\end{align}
	Without aerial platforms, the average number of effective satellites is given by 
	\begin{align}
		&\frac{\lambda\mu }{\pi}\int_0^{\overbar{\xi}} \cos(\varphi)\arcsin\left(\sqrt{1-\cos^2(\overbar{\xi})\sec^2(\varphi)}\right)\diff \varphi.\label{withouta}
	\end{align} 
	Let $\bar{\varphi} = \overbar{\xi}+\overbar{\epsilon} $. For a small $\epsilon$, we show that the aerial platforms increase the average number of effective orbits by a factor of 
	\begin{equation}
		\frac{\sin(\overbar{\varphi})}{\sin (\overbar{\xi})}\approxeq 	\frac{\sin(\overbar{\xi}+\epsilon)}{\sin (\overbar{\xi})}=1+\epsilon\cot(\overbar{\xi}),\label{factororbits}
	\end{equation}
showing that the average number of effective orbits increases linearly with $\epsilon.$ Likewise, aerial platforms increase the effective satellites by a factor of 
	\begin{equation}
		\dfrac{\displaystyle\int_0^{\overbar{\varphi}} \cos(\varphi){\arcsin\left(\sqrt{1-\cos^2(\overbar{\varphi})\sec^2(\varphi)}\right)}\diff \varphi}{\displaystyle\int_0^{\overbar{\xi}} \cos(\varphi){\arcsin\left(\sqrt{1-\cos^2(\overbar{\xi})\sec^2(\varphi)}\right)}\diff \varphi}.\label{factorsatellites}
	\end{equation}
\end{example}
Through system-level simulation and analytical formula in Theorem \ref{Theorem1}, Fig. \ref{fig:theorem1} illustrates the average number of effective satellites as $\lambda$ and $\mu$, the orbit density and the satellite density per orbit, respectively. The system-level simulation was conducted by averaging the network performance over numerous simulation instances where each simulation instance has a single network layout with numerous gateways, based on the proposed network architecture. We confirm that the simulation results confirm the accuracy of the derived formula in Theorem \ref{Theorem1}. The result indicates that aerial platforms effectively increase the number of effective satellites, increasing the likelihood that the typical user  gets messages from satellites. For instance, when $\lambda=15, \mu=10$, only $6$ satellites are available on average; yet, the use of aerial platforms increases this number to $8$ for the same $\lambda$ and $\mu$, exhibiting a $33\%$ increment in the number of effective satellites.

Suppose two scenarios with $\lambda=40, \mu=18$ and $\lambda=27,\mu=27$ having the average numbers of total satellites as $640$ and $729$, respectively. They exhibit a similar average numbers of effective satellites as $30$. Using the result in this paper, we discover that with the help of aerial platform, the same average number of effective satellites is also achieved by a less dense satellite network with $\lambda=27 $ and $\mu=18$ (total $486 $ satellites) This example clearly demonstrates the usefulness of aerial platforms as alternative to dense satellite deployments.

\subsection{Connectivity Probability}
The typical terrestrial gateway  receives the downlink signal if there is at least one  satellite within the extended spherical cap, illustrated as ${\bS}_{A,|\overbar{AC}|} $ in Fig. \ref{fig:fig1concept}.  In the following, we derive the connectivity probability, namely the probability that the typical gateway has at least one  satellite within its extended spherical cap. 
\begin{theorem}\label{Theorem:2}
With aerial platform, the connectivity probability of the typical gateway is given by 
\begin{equation}
	 1- e^{ -\lambda\int_{0}^{\overbar{\varphi}}{\cos(\varphi)}\left(1-e^{-\frac{\mu}{\pi}\arcsin(\sqrt{1-\cos^2(\overbar{\varphi})\sec^2(\varphi)} )}\right)\diff \varphi }. \label{eq:theorem2}
\end{equation}
\end{theorem}
\begin{IEEEproof}
See Appendix \ref{A:2}.
\end{IEEEproof}
\begin{figure}
	\centering
	\includegraphics[width=1.0\linewidth]{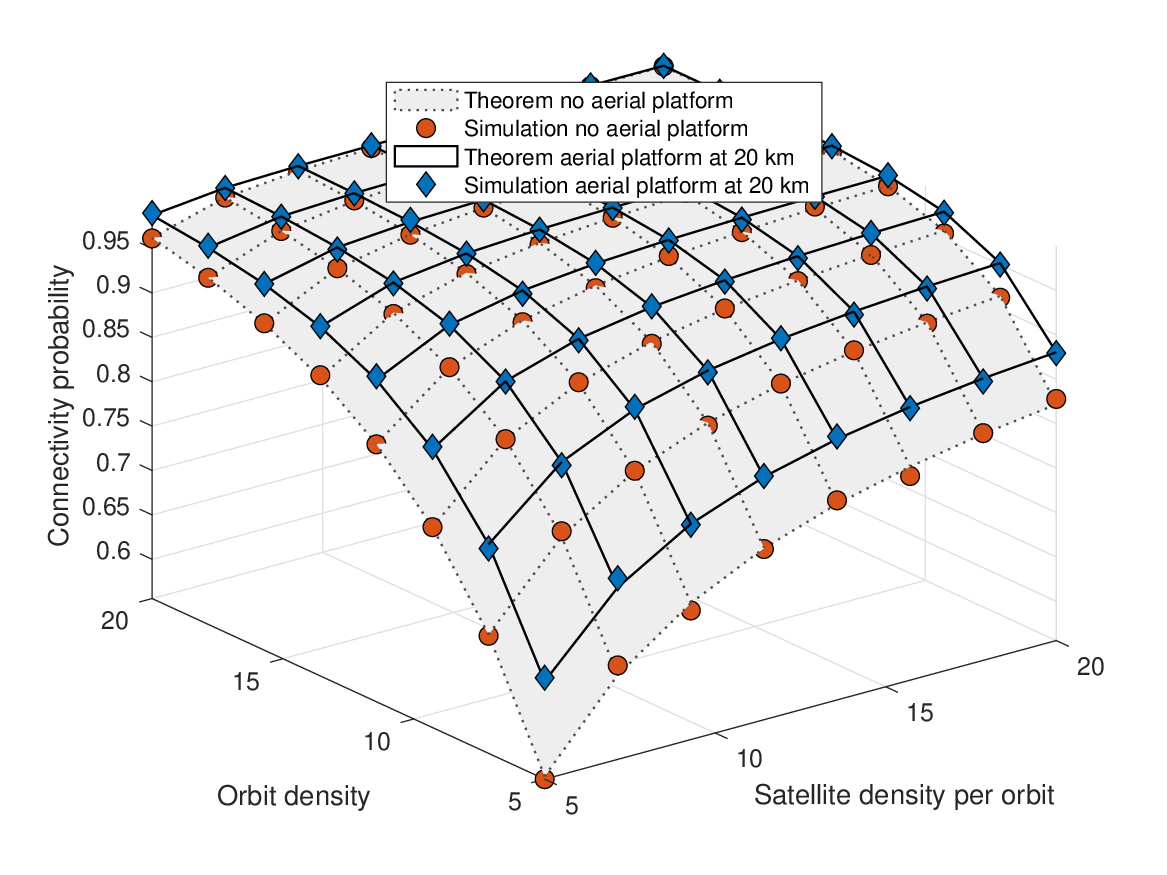}
	\caption{The connectivity probability with and without aerial platforms. The aerial platforms are located at the altitude of $20$ km.}
	\label{fig:theorem2}
\end{figure}

Thanks to the rotation-invariant property we established in proposed network model, the above connectivity probability of the typical gateway corresponds to spatial average of connectivity behavior of \emph{all} users; the number of terrestrial nodes having any effective satellites, divided by the total number of ground nodes. 

Furthermore, because of the time invariant property of the proposed network architecture, the connectivity probability corresponds to the fraction of time that the typical terrestrial gateway has any satellites to connect with; the time slots that the typical gateway has satellites to communicate with, divided by the total time slots over a very long time.  For instance,  the connectivity probability of $0.90$ indicates that the typical gateway has satellites to connect for $90$-time units out of total $100$-time units on average, in the long run.

Fig. \ref{fig:theorem2} displays the connectivity probability of the proposed network model. The simulation results closely align with the synthetic connectivity probability based on Theorem \ref{Theorem:2}, affirming the accuracy of the derived formula. Notably, for low densities of orbits or satellites, the benefits of employing aerial platforms become more evident. In a scenario with sparse satellites, terrestrial nodes struggle to receive signals from satellites because they can get message only from the visible ones. In such cases, introducing aerial platforms atop such terrestrial nodes effectively increases the likelihood of establishing the downlink communications from satellites to such ground nodes.

In Fig. \ref{fig:theorem2}, we observe a connectivity probability of $0.9$ for $\lambda=9,\mu=15$ without platforms, and the same probability is achieved for a less dense satellite network of $\lambda=9,\mu=9$ with aerial platforms. Similarly, without platforms, the connectivity probability for $\lambda=9,\mu=9$ is $0.85$, and the same value is attained with a less dense satellite network of $\lambda=9,\mu=7$ with aerial platforms. These observations suggest that with the assistance of aerial platforms, ground nodes of a sparse  satellite network may have the same or even better connectivity probability, as compared to the ground nodes of a dense  satellite network. Provided that such an increment is observable even in the proposed isotropic satellite distribution, the promised performance gain will appear for various satellite distributions. In Section \ref{S:incomplete}, we show the connectivity gain exists for an existing satellite deployment.

\begin{figure}
	\centering
	\includegraphics[width=1.0\linewidth]{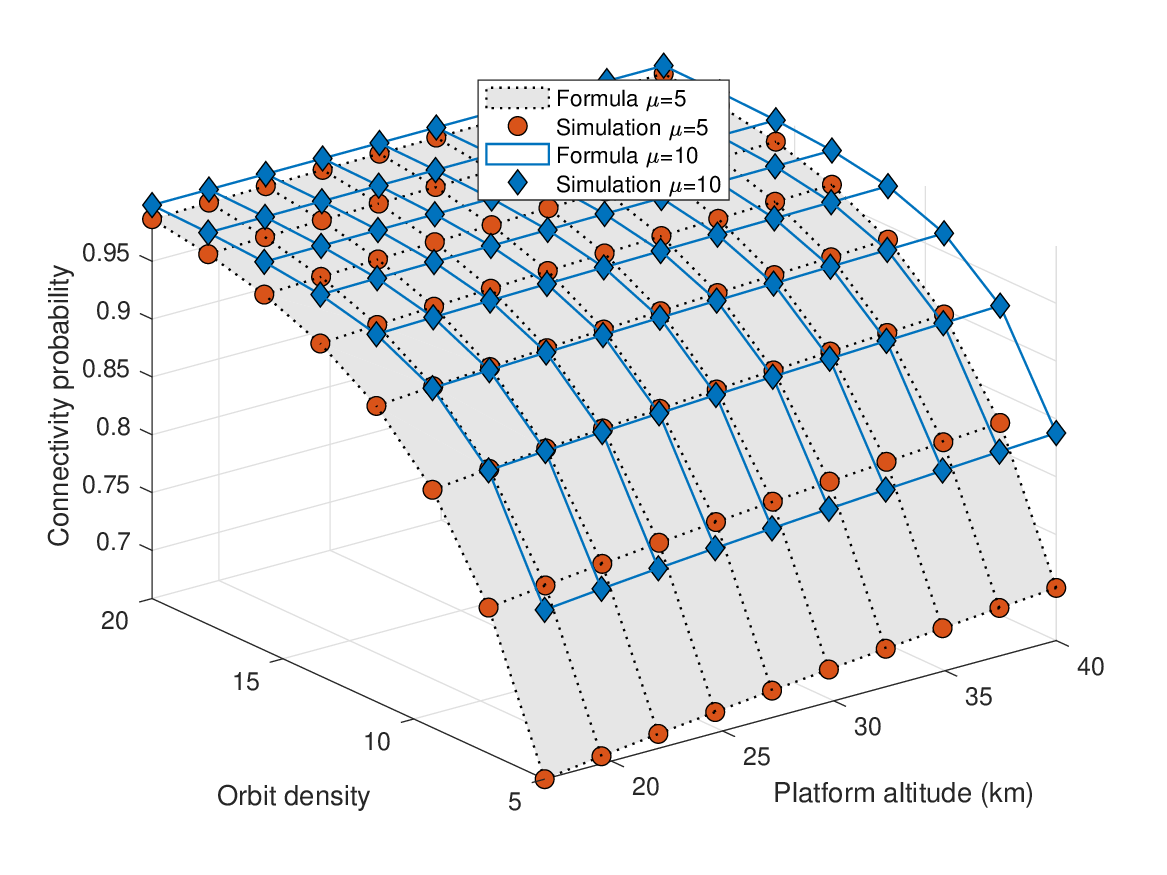}
	\caption{{Connectivity probability of the typical ground gateway. We illustrate two scenarios with $\mu=5$ and $\mu=10$, respectively.}}
	\label{fig:remark1}
\end{figure}

	It is important to note that in Fig. \ref{fig:remark1} the connectivity probability increases with the aerial platform altitude. We confirm that this behavior because---as stated previously and plainly demonstrated in Fig. \ref{fig:fig1concept}---the aerial platforms basically enhance the network connectivity by overcoming the geometric limitation due to the Earth curvature.  

\begin{example}
Without aerial platforms, the connectivity probability is given by the probability that the typical terrestrial gateway has at least one satellite within its visible spherical cap $\bS_{U,|\overbar{UD}|}$ (Fig \ref{fig:fig1concept}). Without aerial platforms, the connectivity probability is 
	\begin{equation}
		1- e^{ -\lambda\int_{0}^{\overbar{\xi}}{\cos(\varphi)}\left(1-e^{-\frac{\mu}{\pi}\arcsin(\sqrt{1-\cos^2(\overbar{\xi})\sec^2(\varphi)} )}\right)\diff \varphi },\label{166a}
	\end{equation}
	where $\overbar{\xi}= \arccos(r_e/r_s).$  By comparing Eqs. \eqref{eq:theorem2} and \eqref{166a}, we show that the aerial platforms  increases the connectivity probability by a multiplicative constant of 
	\begin{equation}
		\frac{1- e^{ -\lambda\int_{0}^{\overbar{\varphi}}{\cos(\varphi)}\left(1-e^{-\frac{\mu}{\pi}\arcsin(\sqrt{1-\cos^2(\overbar{\varphi})\sec^2(\varphi)} )}\right)\diff \varphi }}{1- e^{ -\lambda\int_{0}^{\overbar{\xi}}{\cos(\varphi)}\left(1-e^{-\frac{\mu}{\pi}\arcsin(\sqrt{1-\cos^2(\overbar{\xi})\sec^2(\varphi)} )}\right)\diff \varphi }}.
	\end{equation}
\end{example}

\subsection{Range Distribution}\label{S:Range}
At any given time, for the best signal quality, we assumed that the typical aerial platform is associated with its nearest satellite in the extended spherical cap. This section evaluates the distance distribution from the typical platform to its nearest satellite. 

\begin{theorem}\label{Theorem:3}
	Let us denote by $D$ the distance from the typical aerial platform to its nearest satellite. For $d\leq r_s-r_a, $ $\bP(D>d)  = 1$. For $r_s-r_a< d\leq \sqrt{r_s^2+r_a^2-2r_sr_a\cos(\overbar{\varphi})}$,  $\bP(D>d)$ is given by 
	\begin{align}
		e^{\left(-{\lambda}\int_{0}^{\zeta(d)}\cos(\varphi)\left(1-e^{-\frac{\mu}{\pi}\arcsin\left(\sqrt{1-\cos^2(\zeta(d))\sec^2(\varphi)}\right)} \right)\diff \varphi\right)},\label{eq:theorem3}
	\end{align} 
	where $\zeta(d)= \arccos((r_s^2+r_a^2-d^2)/(2 r_s r_a)).$ For $d>\sqrt{r_s^2+r_a^2-2r_sr_a\cos(\overbar{\varphi})} $, $\bP(D>d)$ is given by 
	\begin{equation}
e^{-\lambda\int_{0}^{\overbar{\varphi}}{\cos(\varphi)}\left(1-e^{-\frac{\mu}{\pi}\arccos(\sqrt{1-\cos^2(\overbar{\varphi})\sec^2(\varphi)} )}\right)\diff \varphi}.\label{eq:theorem3-2}
	\end{equation}
\end{theorem}
\begin{IEEEproof}
See Appendix \ref{A:3}
 	\end{IEEEproof}

\begin{figure}
	\centering
	\includegraphics[width=1.0\linewidth]{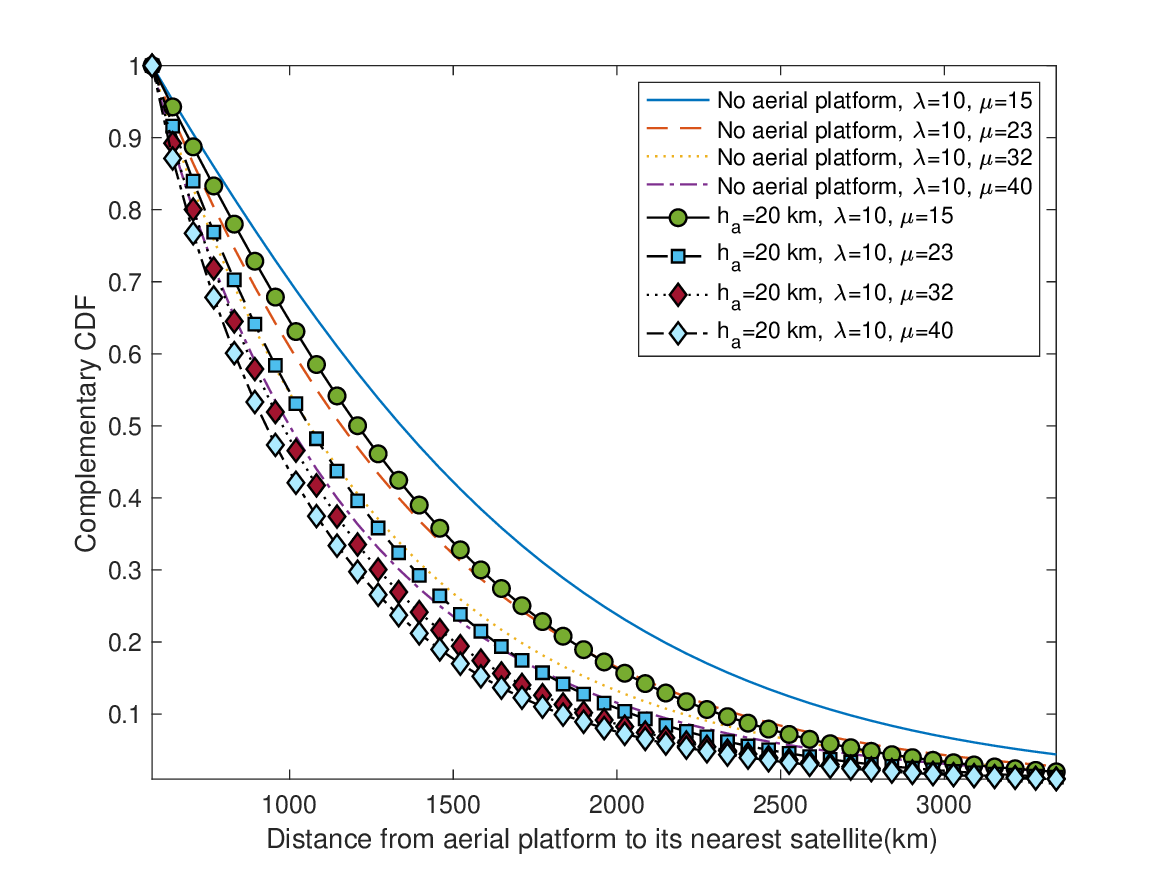}
	\caption{Illustration of the distance distribution from the typical aerial platform to its nearest satellite.}
	\label{fig:theorem3}
\end{figure}
Fig. \ref{fig:theorem3} illustrates the distance from the typical aerial platform to its nearest satellite. In the figure, we fix $\lambda$  at $10$ while varying $\mu$, the satellite density on each orbit from $15$ to $40$. This variation of $\mu$ demonstrates a more conventional strategy of increasing the number of satellites in sparse LEO satellite networks since it is typical for sparse satellite networks to employ a smaller number of orbital planes to densify the network.  As a result, the approach shown in Fig. \ref{fig:theorem3} allows us to investigate and analyze the behavior of the range distribution in response to satellite densification.

In addition, by leveraging the CCDF given in Fig. \ref{fig:theorem3}, we observe that the median values of the random variable $D$ for different scenarios are found at the points where the curves intersect $y=0.5$. We demonstrate that the median distance improves with the addition of aerial platforms to the system. Notably, the improvement in median distance is more pronounced in sparse satellite networks. 

In addition, the median distances remain consistent for a scenario having aerial platforms with $\lambda=10, \mu=15$ and for a scenario having no aerial platform with $\lambda=10, \mu=23$. This implies that utilizing aerial platforms can enable terrestrial gateways to reach median distances comparable to those typically achieved only by a satellite network with $50\%$ more number of  satellites. The significant decrease in median association distance and the comparable median distances achieved by a less dense satellite networks clearly underscore the utility of aerial platforms in sparse  satellite networks, demonstrating their potential as an excellent alternative to dense  satellite networks.

\subsection{SNR Coverage Probability}\label{S:SNRC}

The aerial platforms receive message from  satellites and then they forward the messages to the corresponding ground nodes. In this section, we evaluate the downlink communication performance by evaluating the instantaneous SNR of typical aerial platform and of the typical ground node, respectively. Let $p_s$ and $p_a$ be the received signal power at $1$ meters from satellite transmitters and from aerial platform transmitters, respectively.  Let $B_s$ and $B_a$ be the bandwidth for satellite-to-platform links and for platform-to-terrestrial gateway links, respectively. 


\begin{table}\caption{Simulation parameters}\label{Table:1}
	\centering 
	\begin{tabular}{|c|c|}
		\hline
		Network parameters & values \\
		\hline
		$1$ m average received signal power $p_s $ and $p_g$& $30$ dBm  \\
		\hline
		Average no. of orbits $\lambda$& $25$  \\
		\hline
		Average no. of satellites per orbit $\mu$& $25$  \\
		\hline       
		Satellite altitude $r_s$ & $ 550 $ km \\
		\hline
		Aerial platform altitude $h_a$& $ 20 $ km \\
		\hline		
		Bandwidth $ B_a $ and $ B_g$ & $10$ MHz  \\
		\hline
		Aggregate antenna gain $ g_s g_a $, $g_a g_g$, and $g_s g_g$ &  $26$ dB \\
		\hline
		Carrier frequency $f_c$& $1 $ GHz \\
		\hline
		Noise $N_o$ & $-174 $ dBm/Hz \\
		\hline
		Path loss exponent $\alpha$& $2 $ \\
		\hline
		Satellite angular speed $\nu$ & $0.0011$ rad/s \\ 
		\hline
	\end{tabular}
\end{table}

\begin{theorem}\label{Theorem4}
	The SNR coverage probability of the typical aerial platform is given by Eq. \eqref{eq:theorem4} where we have $|\overbar{AC}| = \sqrt{r_s^2+r_a^2-2r_sr_a\cos(\overbar{\varphi})}$ and  $\overbar{\varphi} = \arccos(r_e/r_a)+\arccos(r_e/r_s)$. We have  $\eta_s = (p_sg_ag_s)/(N_oB_s)$ and $\zeta(z) =\arccos((r_s^2+r_a^2-z^2)/(2 r_s r_a))$. Here, ${\overbar{F}_H}(x)$  is the CCDF of the random variable of $H$ in Eq. \eqref{fading}. 	The SNR coverage probability of the typical terrestrial gateway is 
\begin{equation}
	{\overbar{F}_H}\left({\tau (r_a-r_e)^\alpha}/\eta_a\right),\label{eq:theorem4-1}
\end{equation}
where  $\eta_a = (p_ag_ag_g)/(N_oB_a).$
\end{theorem}

\begin{IEEEproof}
See Appendix \ref{A:4}.
\end{IEEEproof}
Figs. \ref{fig:theorem4} and \ref{fig:theorem4_2} show the SNR coverage probability of the typical aerial platform. By comparing the simulation-based SNR coverage probability with the formula-based SNR coverage probability that we just derived, Fig.  \ref{fig:theorem4} confirms the accuracy of the formula in Theorem \ref{fig:theorem4}.

To obtain the SNR coverage probability through simulation, one must perform system-level simulations where, for each network snapshot, all network nodes are deployed, and the corresponding signal-to-noise ratio for all links are numerically derived and empirically averaged. Therefore, a very large number of simulation instances are required to obtain smooth results. On the other hand, the derived formula computes the coverage probability using a single integral expression, providing results in a much shorter time. This efficiency enables network operators to evaluate network performance by adjusting various parameters simultaneously or individually.

Fig. \ref{fig:theorem4_2} shows the behavior of the SNR coverage probability with respect to $\lambda$ and $\mu$. We observe that increasing $\lambda$ and $\mu$ improves the coverage probability since both parameters result in a shorter communication distance to the associated satellite. Similarly, the presence of an aerial platform also improves the SNR coverage probability.

\begin{figure}
	\centering
	\includegraphics[width=1.0\linewidth]{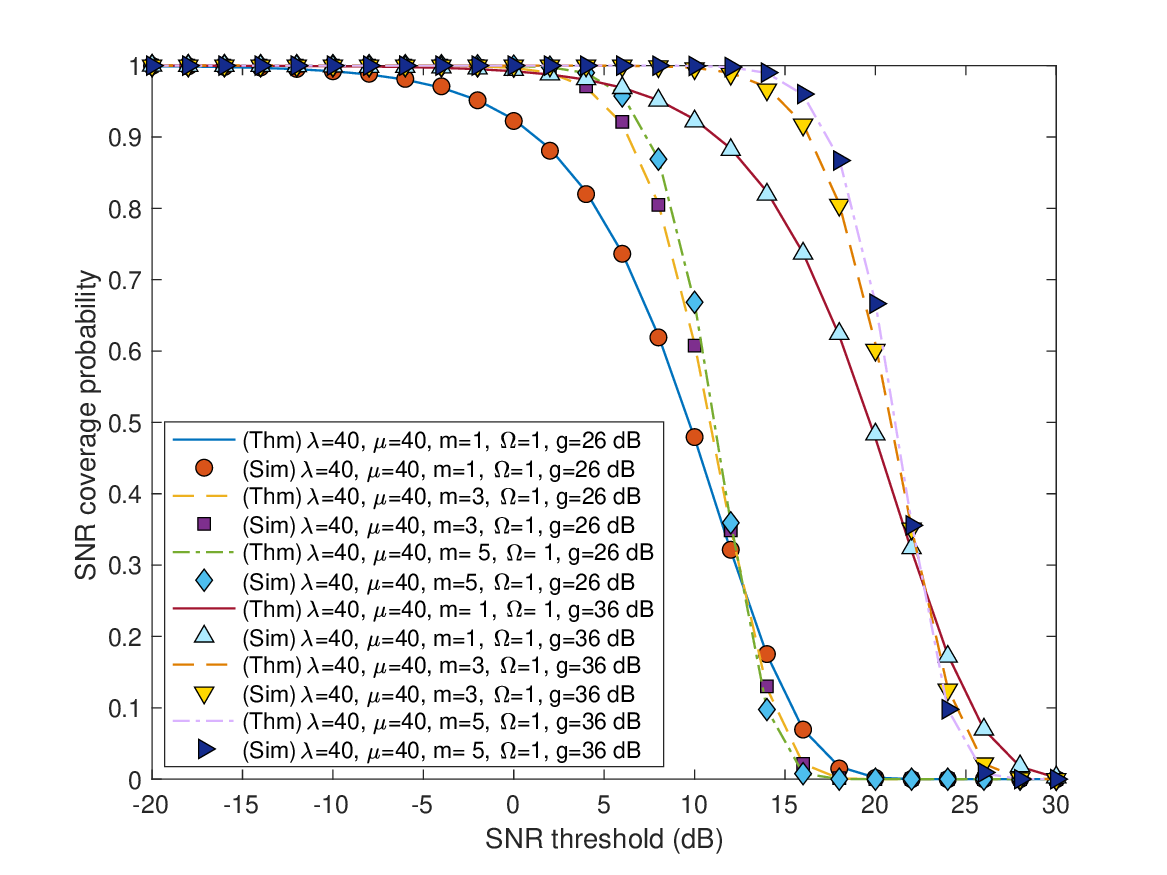}
	\caption{SNR coverage probability of the typical aerial platform. We use parameter values in Table \ref{Table:1}. }
	\label{fig:theorem4}
\end{figure}

\begin{figure}
	\centering
	\includegraphics[width=1.0\linewidth]{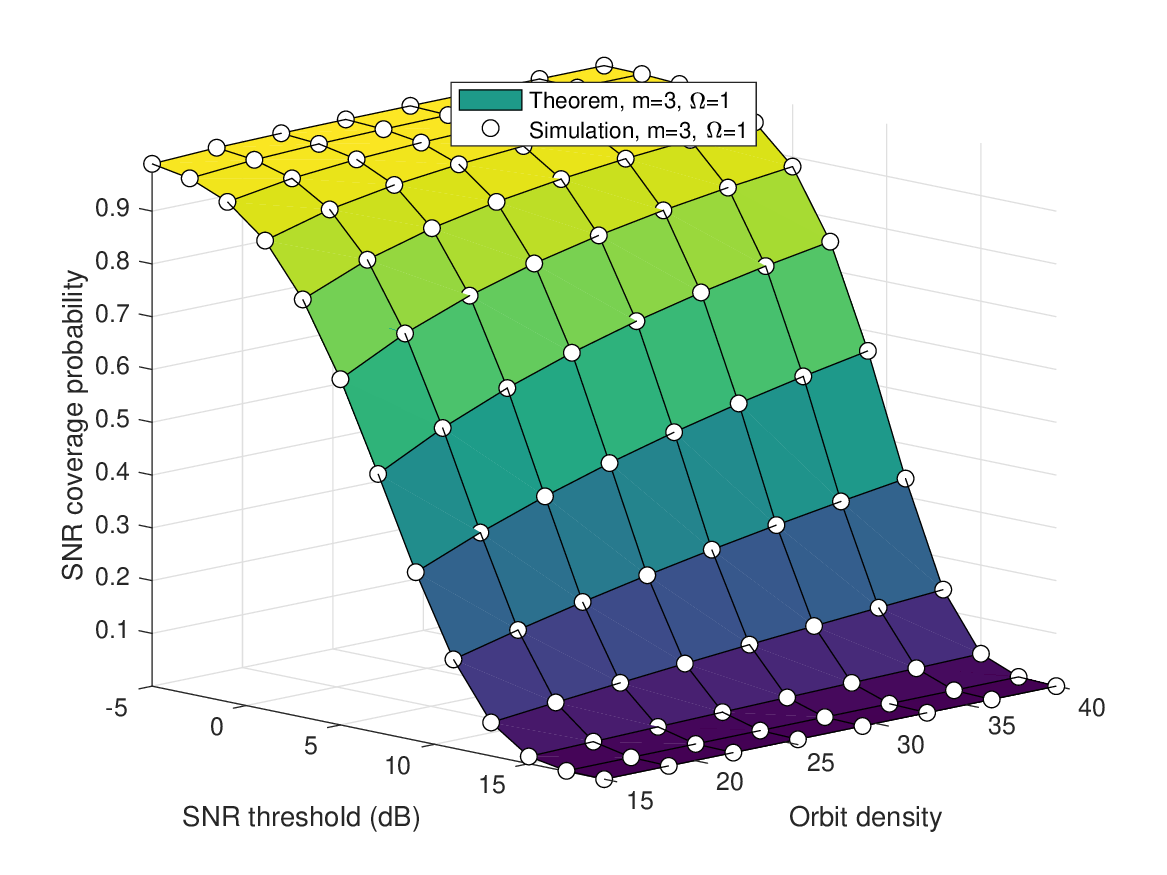}
	\caption{SNR coverage probability of the typical aerial platform. We use the parameter values in Table \ref{Table:1} and $\mu=15$. }
	\label{fig:theorem4_2}
\end{figure}
\begin{figure}
	\centering
	\includegraphics[width=1.0\linewidth]{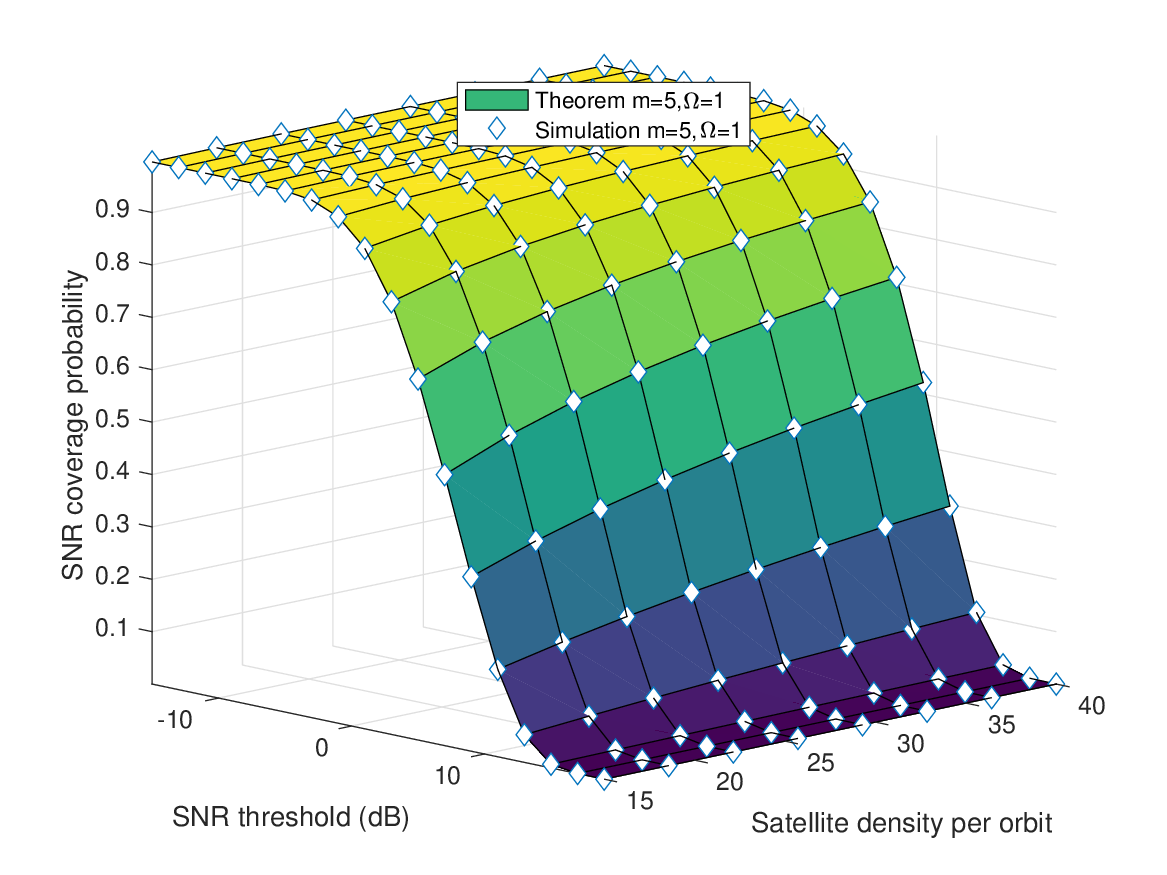}
	\caption{SNR coverage probability of the typical aerial platform. We use the parameter values in Table \ref{Table:1} and $\mu=15$. }
	\label{fig:theorem4_3}
\end{figure}

\begin{figure*}
	\begin{align}
		&\bigintsss_{r_s-r_a}^{|\overbar{AC}|}\bigintsss_{0}^{\zeta(z)}\frac{\lambda\mu z{\overbar{F}_H}\left(\frac{\tau z^\alpha}{\eta_s}\right) e^{-\frac{\mu}{\pi}\arcsin\left(\sqrt{1-\frac{\cos^2(\zeta(z))}{\cos^2(v)}}\right)-{\lambda} \int_{0}^{\zeta(z)} \cos(w)\left( 1 - e^{-\frac{\mu}{\pi}\arcsin(\sqrt{1-\cos^2(\zeta(z))\sec^2(w)})}  \right) \diff w} }{\pi r_s r_a \sqrt{1-\cos^2(\zeta(z))\sec^2(v)}}	  \diff v\diff z.\label{eq:theorem4}\\
				&	\bigintsss_{0}^{\infty} \!\! \bigintsss_{r_s-r_a}^{|\overbar{AC}|} \!\!\bigintsss_{0}^{\zeta(z)}\!\frac{\lambda\mu z{\overbar{F}_H}\left(\frac{(2^u-1) z^\alpha}{\eta_s}\right) e^{-\frac{\mu}{\pi}\arcsin\left(\sqrt{1-\frac{\cos^2(\zeta(z))}{\cos^2(v)}}\right)-{\lambda} \int_{0}^{\zeta(z)} \cos(w) \left(1 - e^{-\frac{\mu}{\pi}\arcsin\left(\sqrt{1-\frac{\cos^2(\zeta(z))}{\cos^2(w)} }\right) }  \right) \diff w}}{\pi r_s r_a \sqrt{1-{\cos^2(\zeta(z))}{\sec^2(v)}}}    \diff v \diff z\diff u.\label{eq:RA}
	\end{align}
	\rule{\linewidth}{0.1mm}
\end{figure*}

\begin{example}\label{example1}
Since the SNR coverage probability of the typical aerial platform is generally represented as a function of the CCDF of the channel random variable $H$, we easily obtain the SNR coverage probability under any fading scenarios by using the CCDF of the corresponding fading random variable. For the shadowed Ricean fading \cite{1198102,6676775}, we have 
\begin{align}
	F_H(x) = K \sum_{n=0}^{\infty} \frac{(\tilde{m})_n \tilde{\delta}^n (2b)^{1+n}}{(n!)^2}\gamma\left(1+n, \frac{x}{2b}\right),\label{eq:fadingothers}
\end{align}
where $K=(2b\tilde{m}/(2b\tilde{m}+\tilde{\Omega}))^{\tilde{m}}/(2b)$ and $\tilde{\delta} = (\tilde{\Omega}/(2b\tilde{m}+\tilde{\Omega}))/2b$. Here, $\tilde{\Omega}$ is the average signal power of the LOS component, $2b$ is the average of powers of the multi-path components except the LOS, and $\tilde{m}$ is the Nakagami parameter. Employing Eqs. \eqref{eq:theorem4} and \eqref{eq:fadingothers}, one can obtain the SNR of the typical aerial platform derived under the shadowed Ricean fading scenario.
\end{example}
\subsection{End-to-End Throughput}
In the proposed network architecture, the end-to-end communications from satellites to aerial platforms are connected via aerial platforms. In this section, we examine the achievable rate of such end-to-end communications. First, the achievable rates of the two links---namely satellite-to-platform and platform-to-ground---are individually derived using Theorem \ref{Theorem4}. Then, we define the end-to-end throughput by taking the minimum of the achievable rates of those two links.

\begin{theorem}\label{Theorem5}
	In the proposed network architecture, the network capacity is given by $\min({R}_A,R_G)$ where $R_A$ is given by \eqref{eq:RA} and $R_G$ is given by 
	\begin{equation}
		\int_0^\infty\overbar{F}_H\left(\frac{(2^u-1) h_a^\alpha}{\eta_g}\right) \diff u,\nnb
	\end{equation}
	respectively.
\end{theorem}
\begin{IEEEproof}
See Appendix \ref{A:5}.
\end{IEEEproof}

Fig. \ref{fig:theorem5} illustrates the network capacity of the proposed network architecture.  First, we observe that the network capacity produced by Theorem \ref{Theorem5} aligns with the simulated network capacity. Secondly, Fig. \ref{fig:theorem5} compare the network capacity with and without aerial platforms. The side-by-side illustration of the plots of these two cases clearly demonstrates that by adding aerial platforms to the system, we have an almost $50\%$ improvement of the network capacity. The inclusion of aerial platforms leads to a shorter communication distance, a better connectivity probability, and a slightly enhanced SNR coverage probability, collectively contributing to the better network capacity, as expected by Theorem \ref{Theorem5} and demonstrated by Fig. \ref{fig:theorem5}. As we discovered in Theorems \ref{Theorem1} -- \ref{Theorem4}, we here also observe the use of aerial platforms gives a comparable or better network performance, even with a less number of  satellites.

\begin{figure}
	\centering
	\includegraphics[width=1.0\linewidth]{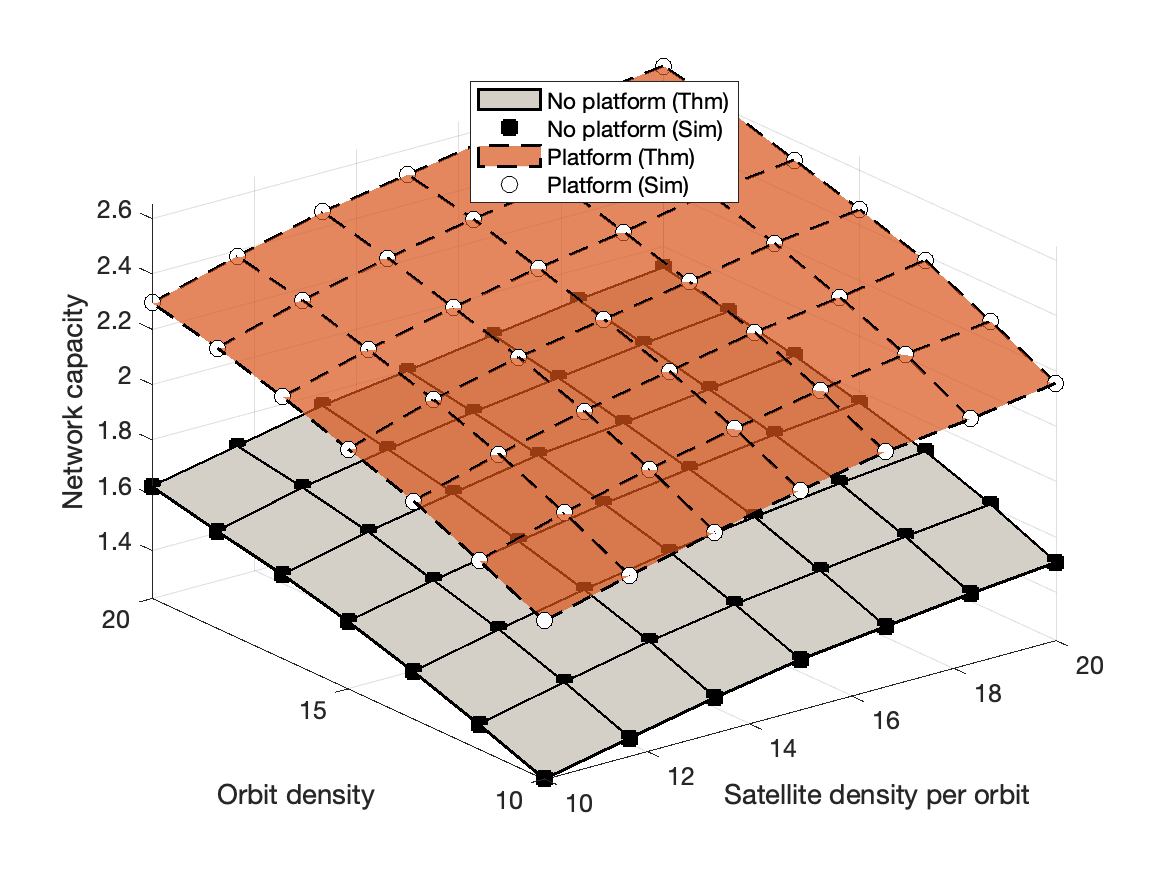}
	\caption{We use values in Table \ref{Table:1}, $N_oB_g=-101$ dBm with $m=1$. Simulation results confirms the analytical formula in Theorem \ref{Theorem5}. }
	\label{fig:theorem5}
\end{figure}

\subsection{Association Delay}
For a sparse  satellite network, ground nodes may need to wait for satellites since it can receive message only from visible satellites\footnote{In the cases where the communication range of satellites is more restrictive, the geometric association delays without aerial platform would be much longer and hence, improvements through aerial platforms will be much greater.  A detailed analysis on this setting is left for future work.}. We define the typical delay as the time that the typical terrestrial gateway is required to wait until any satellites appear. Therefore it is important to note that the delay here is determined solely by the network geometry and independent of the implementation at transmitters and receivers, showing the generality of the investigated metric. With the help of aerial platforms, the typical terrestrial gateway communicates not only with the visible satellites but also the ones invisible yet in the extended spherical cap. Note this paper focuses on the association delay based on network geometry between satellites, aerial platforms, and ground node; other delay aspects including transmission delay or round trip delay are left for future work.

\begin{theorem}\label{Theorem:6}
	The CCDF of the association delay of the proposed network $\bP(T>t)$ is 
	\begin{align}
e^{-{\lambda}\int_{0}^{\overbar{\varphi}}\cos(\varphi)\left(1-e^{-\frac{\mu }{2\pi}\left(\nu t +2\arcsin\left(\sqrt{1-\cos^2(\overbar{\varphi})\csc^2(\phi)}\right)\right)}\right) \diff \varphi },
	\end{align}
	where $\nu$ is the angular speed of the  satellites. 
	\end{theorem}
	\begin{IEEEproof}
		See Appendix \ref{A:6}.
	\end{IEEEproof}

	\begin{figure}
		\centering
		\includegraphics[width=1.0\linewidth]{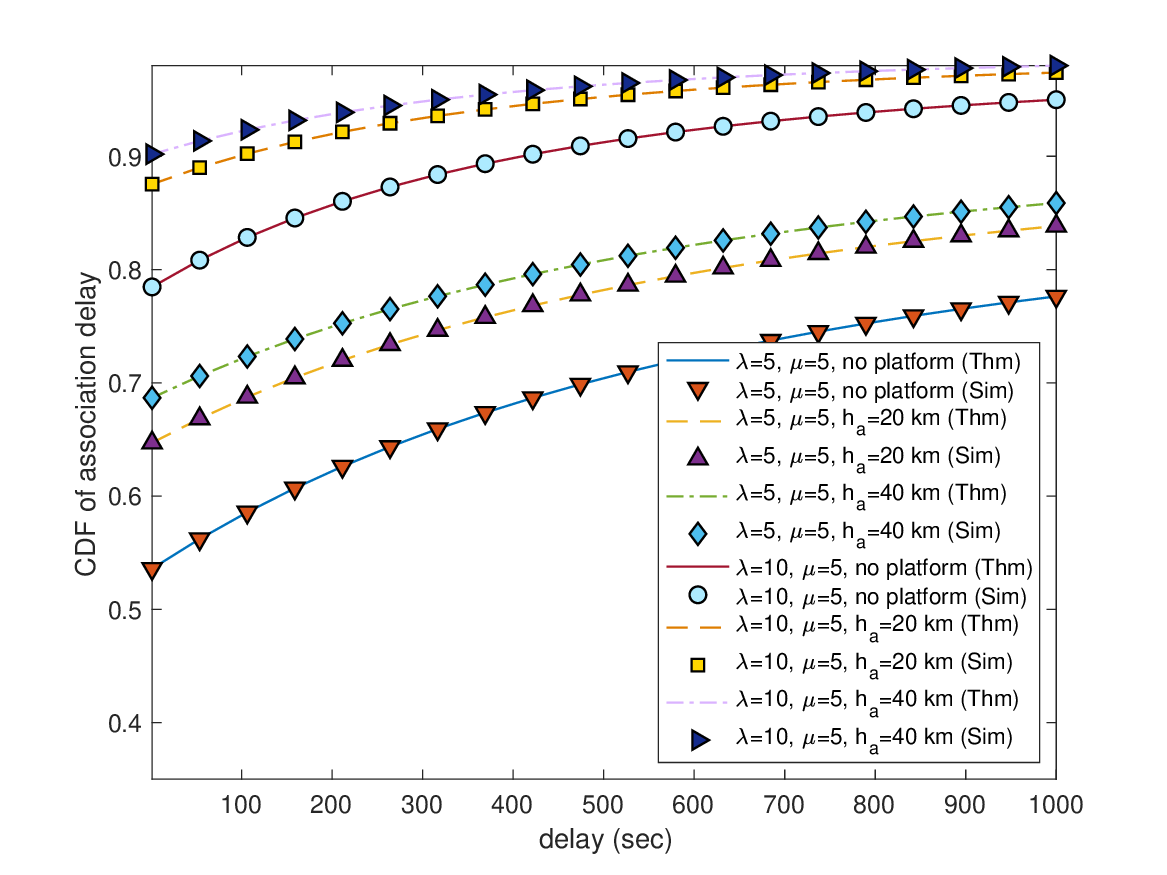}
		\caption{The association delay of the proposed network architecture. The simulation results confirm the derived Theorem \ref{Theorem5}. }
		\label{fig:theorem6}
	\end{figure}
	
Fig. \ref{fig:theorem6} displays the geometric association delay distribution of the proposed network. We observe that aerial platforms significantly reduce the delay. For instance, without aerial platforms, the $70$th percentile delay is about $520$ seconds, whereas with aerial platforms at $h_a=20$ km and $h_a=40$ km, the $70$th percentile delays decrease to $150$ seconds and $50$ seconds, respectively. This indicates that aerial platforms significantly reduce delay in the network architecture, overcoming fundamental limitations imposed by the network geometry, especially for the sparse satellite networks. As consistently demonstrated in Theorems \ref{Theorem1} to \ref{Theorem5}, the results on geometric association delay indicate that aerial platforms in sparse satellite networks serve as a viable alternative to dense satellite networks.

\begin{remark}
In general, there will be other factors determining the total system delay. We analyzed the geometric association delay first because, unlike other factors, the geometric association delay is purely determined by the network geometry and thus independent of implementation. In particular, if the geometric association delay is non-zero, it basically serves as a lower bound on the total system delay.
	
Nevertheless, our framework expands to incorporate other factors, such as propagation delay, given that the geometric association delay is assumed to be zero. For this case, there must  be satellites in the extended spherical cap.
First, the distance from the typical aerial platform to the association satellite is distributed according to the density function we derived in Theorem \eqref{Theorem:3}. Secondly, the distance from the typical aerial platform to the typical terrestrial gateway is the altitude of the aerial platform $r_a-r_e.$ As a result, the total propagation delay is given by Eqs. \eqref{eq:theorem3} divided by the speed of light $c_l=300,000$ km/sec, added to the constant $(r_a-r_e)/c_l$.
\end{remark}
\begin{remark}
For the network operator of a sparse satellite network, it may be possible to deploy more and more satellites into the existing orbits by increasing the number of satellites on each orbit, as we discussed in Sections \ref{S:Range} and \ref{S:SNRC}. Likewise, we can also analyze the time-domain network performance in this asymptotic scenario using Theorem \ref{Theorem:6}. For instance, for a very high $\mu$, the typical association delay distribution of the network model is approximated as follows: 
\begin{align}
	&e^{-{\lambda}\int_{0}^{\overbar{\varphi}}\cos(\varphi)\left(1-e^{-\frac{\mu }{2\pi}\left(\nu t +2\arcsin\left(\sqrt{1-\cos^2(\overbar{\varphi})\csc^2(\phi)}\right)\right)}\right) \diff \varphi }\nnb\\
	&\stackrel{\mu\to \infty}{\approxeq} e^{-{\lambda}\int_{0}^{\overbar{\varphi}}\cos(\varphi)\diff \varphi}=e^{-\lambda \sin(\overbar{\varphi})},\label{eq:Remark2}
\end{align}
where $\overbar{\varphi}$ is given by Eq. \eqref{eq:over_varphi}. This asymptotic result combined with Theorem \ref{Theorem:6} suggests that one can first improve the typical delay of the network by continuously adding more satellites. However, the delay is ultimately upper bounded by the fundamental network geometry such as the number of orbits and the altitude of the aerial platforms. To reduce the delay further, one must increase either $\lambda$ the mean number of orbits, or $r_a$ the aerial platform altitude, or $r_s$ the satellite radius. The above asymptotic analysis indicates how much gain is achievable in terms of typical delay when each orbit is densely populated with enough satellites.
\end{remark}
\section{Discussion}\label{Section:Discussion}
In this section, we discuss the further development of the proposed network architecture to address some practical limitations. We also discuss potential future work based on the proposed framework. 

\subsection{Random Elevation Angle of Aerial Platform}\label{S:elevation}

This paper assumes that the typical aerial platform is positioned directly above the typical terrestrial node. In terms of establishing secure communications, this higher elevation angle is preferred, as it increases the likelihood of LOS communications, enhances signal strength, and reduces atmospheric effects such as ground refraction or scattering.

Below, we expand our analysis by assuming a random elevation angle for the aerial platform to further assess its impact. To evaluate network performance, we utilize the stochastic geometry framework. For this analysis, let us define the zenith angle $Z$ as $90$ minus the elevation angle of the typical aerial platform.

\begin{proposition}\label{Proposition:1}
Suppose a random variable $Z$ for the zenith angle with its PDF being $f_Z(z)$ and its CDF being $ F_Z(z), $  respectively. The connectivity probability with aerial platforms is given by 
		\begin{equation*}
			1-\bE\left[e^{-\lambda\int_{0}^{\hat{\varphi}}{\cos(\varphi)}\left(1-e^{-\frac{\mu}{\pi}\arcsin\left(\sqrt{1-\cos^2(\hat{\varphi})\sec^2(\varphi)}\right)}\right)\diff \varphi}\right], 
		\end{equation*}
		where $\hat{\varphi}  = \arccos(r_e/r_l)+\arccos(r_e/r_s)$. In above, the expectation is w.r.t $Z$
and $r_l = \sqrt{r_a^2+(r_a-r_e)^2\tan^2(Z)}$. 
 \end{proposition}
\begin{IEEEproof}
	See Appendix \ref{A:7}. 
\end{IEEEproof}
\subsection{Minimum Elevation Angle from Terrestrial Node}

In practice, terrain or uneven Earth surface may obstruct the signals from satellites to terrestrial nodes, particularly when the elevation angles of the satellites are low. Although the main analysis in Section \ref{Section:result} has been conducted without considering the impact of terrain, we analyze the impact of uneven Earth surface in below by assuming that the terrestrial nodes can receive messages only from satellites whose elevation angles are greater than a certain predefined angle $\kappa.$
\begin{proposition}\label{Proposition:2}
	Assuming the minimum elevation angle $\kappa$ from the terrestrial nodes, the connectivity probability without aerial platforms is given by 
			\begin{equation}
			1- e^{ -\lambda\int_{0}^{\tilde{\xi}}{\cos(\varphi)}\left(1-e^{-\frac{\mu}{\pi}\arcsin(\sqrt{1-\cos^2(\overbar{\xi})\sec^2(\varphi)} )}\right)\diff \varphi },\label{28}
		\end{equation}
		where we have
		\begin{equation}
			\tilde{\xi} = \left.\arccos\left(\frac{r_e^2+r_s^2-y^2}{2 r_s r_e}\right)\right|_{y=r_e\sin(\kappa)+\sqrt{r_e^2\cos^2(\kappa)+r_s^2}}. 
		\end{equation}
\end{proposition}
\begin{IEEEproof}
	See Appendix \ref{A:8}. 
\end{IEEEproof}
It is reasonable to assume that the limit on the elevation angle due to an uneven Earth surface does not radically change the connectivity probability of the typical user being assisted by aerial platform. On the other hand, On the other hand, the impact of terrain could be studied further by expanding the approach of Proposition \ref{Proposition:1}, treating $Z$ as a mixed random variable with a finite support. This is left for future work.

 \subsection{Performance Gain for Incomplete Constellation}\label{S:incomplete}
 
First, it is important to note that any existing or forthcoming satellite networks can be accurately represented by the proposed Cox model through a moment-matching method, where the average numbers of target satellites and the Cox-distributed satellites are matched. Using this method, an accurate representation of the network performance of the target constellation is successfully reproduced. Empirical evidence for this was provided in \cite{choi2023, choi2023cox, choi2023delay_COX, choi2023hetero_COX}. 

\begin{figure}
	\centering
	\includegraphics[width=1.0\linewidth]{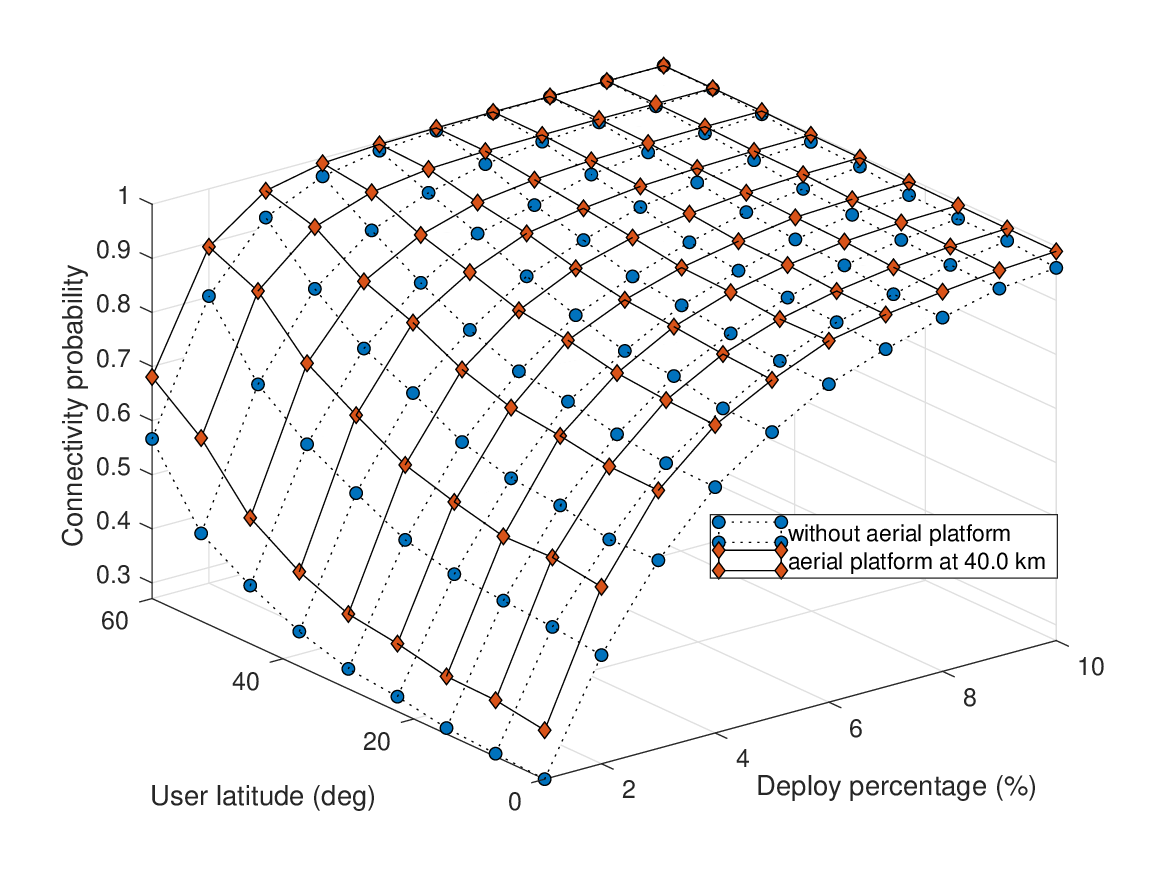}
	\caption{The benefit of employing aerial platform at altitude of $40$ km. For all deployment stage and all user latitude, we observe that aerial platform improves the connectivity probability of the  satellite network.}
	\label{fig:section4cconnectcompare}
\end{figure}

Another interesting application of the proposed model is its ability to represent various development stages of a given constellation and demonstrate the benefits of aerial platforms in terms of overall network performance. Let us consider an existing OneWeb constellation with $600$ satellites. We vary the deployment stage of the OneWeb constellation from $0$\% to $10$\% and evaluate the connectivity probability with and without aerial platforms. As observed in Fig. \ref{fig:section4cconnectcompare}, aerial platforms significantly improve the connectivity probability of the satellite network, expanding the limited coverage during its initial stage. The $x$-axis is the user latitude and the $y$-axis is the deployment percentage, with respect to its complete stage.

\subsection{Operational Challenges}
Here are some specific operational and technical challenges in using aerial platforms at high altitude as relaying agents:
\begin{itemize}
	\item Lightweight structure: Due to thin air density, the wing area of an aerial platform must be large. Therefore, the material must have high strength and durability to withstand any damage \cite{yokomaku2000overview}.
	\item Energy generation and storage: An aerial platform requires a source of energy to enable the propulsion system and power communication equipment. An energy-efficient design was proposed in \cite{nickol2007high}.
	\item Thermal management: An aerial platform is influenced by temperature differences, undergoing superheating and supercooling. Some challenges and technologies were discussed in \cite{wu2015thermal}.
\end{itemize}
For maintaining effective communications with satellites and terrestrial nodes, the above technological challenges need to be addressed.

\subsection{Future Work: Aerial Platform Sharing}

To improve the coverage of a LEO satellite network, this paper assumes that an aerial platform is located directly above each terrestrial node. For a sparse satellite network with a minimum number of satellites, such a deployment strategy is feasible. However, one could expand the current architecture of one-to-one mapping to include a scenario where a single aerial platform serves multiple terrestrial nodes. In this case, the aerial platforms are not necessarily positioned above each terrestrial node, and therefore, the distance from terrestrial nodes to aerial platforms must be evaluated and analyzed.

For the sake of analysis, we may assume that aerial platforms and terrestrial nodes are modeled as binomial point processes on concentric spheres of different radius, with the density of terrestrial nodes being larger than that of aerial platforms. Similarly, since terrestrial nodes share aerial platforms, we also assume that terrestrial nodes are associated with their closest aerial platforms and that each aerial platform shares its time between its associated terrestrial nodes. The distance from the typical terrestrial gateway  to its associated (nearest) aerial platform must be derived, and this analysis can be directly obtained from the literature \cite{9177073}, where the distance from a typical random point to a binomial point process on a sphere was derived.

In this new scenario, the distance is no longer constant but rather a function of the density of aerial platforms. By combining this new association distance for the aerial platform sharing scenario with the provided analysis on the link between a typical aerial platform and satellites, we can obtain the network performance of this new architecture. Since terrestrial nodes now share aerial platforms, the link distances from terrestrial nodes to aerial platforms are strictly smaller than in the non-sharing scenario, and thus, the network gain from the aerial platforms is expected to decrease as well. A more detailed analysis of this potential expansion of aerial platform sharing is left for future work.

\section{Conclusion}
This paper explores the potential of aerial platforms as an excellent alternative to a dense satellite network, which has been considered essential for seamless and reliable communications. We present a space-time 4D model where the motion and locations of satellites in orbits are unified through the conditional structure of the Cox point process. This model allows us to analyze not only spatial domain performance metrics, such as the SNR coverage probability or the number of orbits, but also temporal domain performance metrics, such as the time fraction of connectivity and the association delay. The derived performance metrics are compared to simulation results for accuracy, consistently showing that the addition of aerial platforms enables downlink communication performance that would otherwise require a dense satellite network. This explicitly demonstrates the viability of employing aerial platforms when densely deployed satellites are unavailable.

Future work can further examine the benefits of aerial platforms in more practical settings, such as uneven terrain and restrictive elevation angles. Likewise, by advancing the developed framework, network performance could be analyzed in extended scenarios, such as aerial platform sharing or periodic satellite distribution and motion. Building upon the preliminary results established in this work, future research may explore the use of aerial platforms in satellite networks by assessing the network performance achieved with aerial platforms or demonstrating their applicability as an alternative to dense satellite networks.


\appendices

\section{Proof of Theorem \ref{Theorem1}}\label{A:1} 
	The typical aerial platform is at the distance $r_a$ from the origin, right above the typical ground node. Due to the typical aerial platform assisting the downlink communications, the extended spherical cap is given by  ${\bS}_{A,|\overbar{AC}|}$ in Fig. \ref{fig:fig1concept}. 	Let $\overbar{\varphi}$ be the angle $\angle AOC$. 	We get 
	\begin{align}
		\overbar{\varphi}&= \angle AOB + \angle BOC \nnb\\
		&= \arccos(r_e/r_a)+ \arccos(r_e/r_s).\label{eq:over_varphi}
	\end{align}
	For all values of $r_e, r_a$, and $r_s$, we have $\overbar{\varphi}\leq\pi/2$ and the equality holds iff $r_a, r_s\gg r_e$.

	Using $\overbar{\varphi}$, we obtain $|\overbar{AC}|= \sqrt{r_a^2+r_s^2-2r_s r_a\cos(\overbar{\varphi}) }$. The extended cap $		{\bS}_{A,|\overbar{AC}|}$ is defined as 
	\begin{equation}
		\{(x,y,z)\in \bS_{r_s} | \|(x,y,z) -A \|  \! \leq \! \sqrt{r_a^2+r_s^2-2r_s r_a \cos(\overbar{\varphi})}\}.\nnb
	\end{equation}
	Then, the orbit $l(\theta,\phi)$ meet the extended spherical cap ${\bS}_{A,|\overbar{AC}|}$ if $\phi\in [\pi/2-\overbar{\varphi},\pi/2 + \overbar{\varphi}]$. To derive the number of such orbits in the extended spherical cap, we use the Campbell's mean value theorem to have the following expression.
	\begin{align}
		&\bE\left[\Xi([0,\pi]\times \left[\frac{\pi}{2}-\overbar{\varphi},\frac{\pi}{2}+\overbar{\varphi}\right]\right] \nnb\\
		&= \int_{0}^{\pi}\int_{\pi/2-\overbar{\varphi}}^{\pi/2+\overbar{\varphi}}\frac{\lambda\sin(\phi)}{2\pi}\diff \theta\diff \phi=\lambda\sin(\overbar{\varphi}),
	\end{align}
	where we use the fact that the density of the orbit process $\Xi$ is $\lambda\sin(\phi)/(2\pi)$ on $\cR$.
	\par To derive the average number of effective satellites in the cap $\bS_{A,|\overbar{AC}|}$, we first use the fact that conditionally on orbits, the satellites Poisson point processes over different orbits are independent. Therefore, the effective number of satellites is given by the summation of satellite Poisson point processes as follows:  
	\begin{align}
		\bE[\Psi({\bS}_{A,|\overbar{AC}|})]& = \bE\left[\sum_{i\in\bN} \psi_i({\bS}_{A,|\overbar{AC}|})\right]\nnb\\
		&=\bE\left[\sum_{i\in\bN}\bE\left[\left.\psi_i({\bS}_{A,|\overbar{AC}|})\right| O \right]\right]\nnb\\
		&=\bE\left[\sum_{i\in\bN}^{|\pi/2-\phi_i|<\overbar{\varphi}}\bE\left[\psi_i({\bS}_{A,|\overbar{AC}|})\right]\right],\label{16}
	\end{align} 
	where we again use the fact that only orbits with $\phi\in[\pi/2-\overbar{\varphi},\pi/2+\overbar{\varphi}]$ meet the effective spherical cap ${\bS}_{A,|\overbar{AC}|}$ and the effective  satellites are on the intersection of such orbits and the effective spherical cap. 
	
	Conditionally on the orbit process $\cO$, the satellite points of the orbit $l(\theta_i,\phi_i)$ on the extended spherical cap ${\bS}_{U,|\overbar{UC}|}$ is a Poisson point process of intensity $\mu/(2\pi r_s)$. Therefore, we have 
	\begin{align}
		\bE\left[\psi_i({\bS}_{A,|\overbar{AC}|})\right] &= \frac{\mu}{2\pi r_s}  {|l(\theta_i,\phi_i) \cap  {\bS}_{A,|\overbar{AC}|} |}\nnb\\
		&=\frac{\mu}{\pi} {\arcsin(\sqrt{1-\cos^2(\overbar{\varphi})\csc^2(\phi_i)})}, \label{eq:length}
	\end{align}
	where we use that the length of $l(\theta_i,\phi_i) \cap {\bS}_{A,|\overbar{AC}|}$ is given by $2 r_s \arcsin\left(\sqrt{1-\cos^2(\overbar{\varphi})\csc^2(\phi_i)}\right)$ where $\phi_{i}\in (\pi/2-\overbar{\varphi},\pi/2+\overbar{\varphi})$ in Section \ref{S:prelim}.
	
	Therefore, by combining Eqs. \eqref{16} and \eqref{eq:length}, the average number of effective satellites is 
	\begin{align}
		&\bE[\Psi(\overbar{\bS}_{A,|\overbar{AC}|})]\nnb\\
		&= \bE\left[\sum_{(\theta_i,\phi_i)\in\Xi}^{|\pi/2-\phi_i|<\overbar{\varphi}} \frac{\mu}{\pi} {\arcsin(\sqrt{1-\cos^2(\overbar{\varphi})\csc^2(\phi_i)})} \right]\nnb\\
		&=\int_{\pi/2-\overbar{\varphi}}^{\pi/2+\overbar{\varphi}} \frac{\lambda\mu \sin(\phi)}{2\pi} {\arcsin(\sqrt{1-\cos^2(\overbar{\varphi})\csc^2(\phi)})}\diff \phi\nnb\\
		&=\frac{\lambda\mu }{\pi}\int_0^{\overbar{\varphi}} \cos(\varphi){\arcsin(\sqrt{1-\cos^2(\overbar{\varphi})\sec^2(\varphi)})}\diff \varphi,\label{166}
	\end{align} 
	where we again use Campbell's mean value theorem \cite{baccelli2010stochastic} to integrate the corresponding function over the Poisson point process $\Xi$. Finally, to get Eq. \eqref{166}, we use the change of variables $\varphi = \pi/2-\phi $.\footnote{Note the variable  $\varphi$ can be interpreted as the complement inclination of the orbit since $\varphi = \pi/2-\phi$. As $\phi$ takes values in the interval $[0,\pi),$ the complement inclination $\varphi$ takes values in the interval $ [-\pi/2,\pi/2)$. }

\section{Proof of Theorem \ref{Theorem:2}}\label{A:2}

	The connectivity probability is equal to the probability that there is at least one satellite within the extended spherical cap. Therefore, the connectivity probability is 
\begin{align}
	\bP(\text{connectivity}) &= 1- \bP(\text{no satellite within } {\bS}_{A,|\overbar{AC}|})\nnb\\
	&=1- \bP\left(\forall X_{j,i}\in\psi_i,  \psi_i({\bS}_{A,|\overbar{AC}|})=\emptyset\right)\nnb\\
	&= 1- \bE\left[\prod_{\phi_i\in\Xi}\bE\left[\left.\ind_{\psi_i({\bS}_{A,|\overbar{AC}|})=\emptyset} \right|O\right]\right].\label{25}
\end{align}
Then, by leveraging the fact that the satellite point on the intersection of $l(\theta_i,\phi_i)$ and the cap $\overbar{\bS}_{A,\overbar{AC}} $ is a Poisson point process of intensity $\mu/(2\pi r_s)$, we have 
\begin{align}
	&\bE\left[\left.\ind_{\psi_i({\bS}_{A,|\overbar{AC}|})=\emptyset} \right|O\right]\nnb\\ 
	&= \exp\left(-\frac{\mu}{2\pi r_s}|l(\theta_i,\varphi_i)\cap {\bS}_{A,\overbar{AC}}|\right)\nnb\\
	&=\exp\left(-\frac{\mu}{\pi}\arcsin\left(\sqrt{1-\cos^2(\overbar{\varphi})\csc^2(\phi)} \right)\right), \label{26}
\end{align}
where we use the void probability of the Poisson point process of intensity $\mu/(2\pi r_s)$. 

Moreover, by combining Eqs. \eqref{25} and \eqref{26}, we have 
\begin{align}
	&\bE\left[\prod_{\phi_i\in\Xi}\bE\left[\left.\ind_{\psi_i({\bS}_{A,|\overbar{AC}|})=\emptyset} \right|O\right]\right]\nnb\\
	&=\bE\left[\prod_{\phi_i\in\Xi}^{|\phi_{i}-\pi/2|<\overbar{\varphi}}e^{-\frac{\mu}{\pi}\arcsin\left(\sqrt{1-\cos^2(\overbar{\varphi})\csc^2(\phi_{i})} \right)}\right]\nnb\\
	&=e^{-\lambda\int_{0}^{\overbar{\varphi}}{\cos(\varphi)}\left(1-e^{-\frac{\mu}{\pi}\arcsin\left(\sqrt{1-\cos^2(\overbar{\varphi})\sec^2(\varphi)}\right)}\right)\diff \varphi},\label{27}
\end{align} 
where we first use the probability generating functional of the Poisson point process $\Xi$ of density $\lambda \sin(\phi)/(2\pi)$ and then we employ the change of variable $\varphi = \pi/2-\phi$. Finally, the connectivity probability is given by one minus the final term of Eq. \eqref{27}.   

\section{Proof of Theorem \ref{Theorem:3}}\label{A:3}

	When there is at least one satellite in $\bS_{A,\overbar{AC}}$, the random variable $D$ takes a value within the interval $[r_s-r_a,\sqrt{r_a^2+r_s^2-2r_sr_a\cos(\overbar{\varphi})}]$ where $\overbar{\varphi}$ is given by Eq. \eqref{eq:over_varphi}.  The CCDF of $D$ is  
\begin{align}
	\bP(D>d) &= \bP(|\overbar{AX_{j,i}}|>d, \forall X_{j,i}\in\psi_i,\forall\psi_i\in\Xi )\nnb\\ 
	&\ea\bP\left(\prod_{\phi_i\in\Xi}\bP\left(\psi_i({\bS}_{A,d})= \emptyset\right)  \right)\nnb\\
	&\eb\bP\left(\prod_{\phi_i\in\Xi}e^{-\frac{\mu}{\pi}\arcsin\left(\sqrt{1-\cos^2(\zeta(d))\csc^2(\phi_i)}\right)} \right),\nnb
\end{align}
for $r_s-r_a<d<\sqrt{\|A\|^2+r_s^2-2r_s\|A\|\cos(\zeta(d))}$. To get (a), we use the fact that orbits  are independent and that the satellite Poisson point process on each orbit must have no point on the spherical cap ${\bS}_{A,d}$. To get (b),  we use the expression that the arc length of the intersection of the spherical cap ${\bS}_{A,d}$ and the orbit $l(\theta_i,\phi_i)$ is given by   $2r_s\arcsin\left(\sqrt{1-\cos^2(\zeta(d))\csc^2(\phi_i)}\right)$ where $\zeta(d)$ is 
\begin{align}
	\zeta(d) &= \arccos((r_s^2+r_a^2-d^2)/(2 r_s r_a)).
\end{align}
By employing the probability generating functional of the Poisson point process $\Xi$ of density $\lambda\sin(\phi)/(2\pi )$ on the set $[0,\pi)\times [\pi/2-\zeta(d),\pi/2+\zeta(d)]\subset \cR$, the CCDF of $D$ is 
\begin{align}
	&\bP(D>d)\nnb\\
	& = e^{-\int_{\pi/2-\zeta(d)}^{\pi/2+\zeta(d)}\frac{\lambda\sin(\phi)}{2}\left(1-e^{-\frac{\mu}{\pi}\arcsin\left(\sqrt{1-\cos^2(\zeta(d))\csc^2(\phi)}\right)} \right)\diff \phi }\nnb\\
	&=e^{-{\lambda}\int_{0}^{\zeta(d)}\cos(\varphi)\left(1-e^{-\frac{\mu}{\pi}\arcsin\left(\sqrt{1-\cos^2(\zeta(d))\sec^2(\varphi)}\right)} \right)\diff \varphi},
\end{align}
where we use the change of variable $\phi = \pi/2-\varphi$ to obtain the final result. 
\par On the other hand, for $d>\sqrt{r_s^2+r_a^2-2r_sr_a\cos(\overbar{\varphi})}, $ this case corresponds to the event that there is no  satellite in the extended spherical cap and thus it coincides with  the result of Theorem \ref{Theorem:2}. 

\section{Proof of Theorem \ref{Theorem4}}\label{A:4}

	The SNR coverage probability of the typical aerial platform is given by 
\begin{align}
	&\bP(\SNR_{A}\geq\tau)\nnb\\
	& = \bP\left(\frac{p_sg_sg_aHD^{-\alpha}}{N_oB_s}\geq\tau\right)\nnb\\
	&=\bE\left[\bE\left[\bP\left(\left.\frac{p_sg_sg_aHD^{-\alpha}}{N_oB_s}\geq\tau \right|  l_\star ,\Xi \right)\right]\right]\nnb\\
	&=\bE\left[\bE\left[\bE\left[\bP\left(\left.\frac{p_sg_sg_aHD^{-\alpha}}{N_oB_s}\geq\tau \right| D, l_\star,\Xi\right)\right]\right]\right]\nnb,
\end{align}
where the SNR coverage probability is given by the conditional expectations. In the above expression, $\Xi$ denotes the orbit process, $l_\star$ denotes the orbit that contains the association satellite, and $D$ denotes the distance from the typical platform to the association satellite. 

\par Conditionally on the $D, l_\star, $ and $\Xi$, the probability expression inside the expectation is 
\begin{align}
	&\bP\left(\left.\frac{p_sg_sg_aHD^{-\alpha}}{N_oB_s}>\tau \right| D, l_\star,\Xi\right)\nnb\\
	&=\bP\left(\left.{H}>\frac{\tau N_oB_s D^\alpha}{p_sg_sg_a} \right| D, l_\star,\Xi\right)=\overbar{F}_H\left({\tau D^\alpha}/{\eta_s}\right),\nnb
\end{align}
where $\overbar{F}_H(x)$ is the CCDF of the random variable $H$ evaluated at $x$ and $\eta_s = (p_sg_gg_s)/(N_oB_s).$ Now, conditionally on $\Xi$ and $l_\star,$ the PDF of the distance to the association satellite is 
\begin{align}
	f_D(z) &= \bP(D\in [z,z+\diff z)|l_\star,\Xi) \nnb\\
	&\ea\frac{\diff}{\diff z}\left(1-\bP(\psi_\star(\bS_{A,z})=\emptyset) \right)\prod_{\psi_i\in\Xi}\bP(\psi_i(\bS_{A,z})=\emptyset)\nnb\\
	&\eb\frac{\mu z\csc(\phi_\star)e^{-\frac{\mu}{\pi}\arcsin(\sqrt{1-\cos^2(\zeta(z))\csc^2(\phi_\star)})}}{r_sr_a\sqrt{1-\cos^2(\zeta(z)\csc(\phi_\star))}}\nnb\\
	&\hspace{4mm}\prod_{\psi_i\in\Xi}e^{-\frac{\mu}{\pi}\arcsin(\sqrt{1-\cos^2(\zeta(z))\csc^2(\phi_i)})}\label{eq:45},
\end{align}
where $\psi_\star$ is the satellite Poisson point process of the line $l_\star$ that contains the association satellite $X_\star$. To get (a), we use the fact that $D>z$ iff all the satellites are at distance greater than $z$ from the typical aerial platform. Then, we use  the Slivynak's theorem of the Poisson point process \cite{baccelli2010stochastic}. The SNR coverage probability of the typical aerial platform is given by 
\begin{align}
	&\bP(\SNR_{A}>\tau)	\nnb\\&=\bE_\Xi\left[\bE_{l_\star}\left[\bE_D\left[ \overbar{F}_H(\tau D^\alpha/\eta_s)\right]\right]\right]\nnb\\
	&=\bE_\Xi\left[\bE_{l_\star}\left[  \int_{r_s-r_a}^{|\overbar{AC}|}\overbar{F}_H(\tau D^\alpha/\eta_s)f_D(z)\diff z  \right]\right] \label{40},
\end{align}
where we use Fubini's theorem; namely we integrate with respect to $l_\star$, then $\Xi$, and finally $D$. Integrating w.r.t. the random variable of $l_\star$ gives the following expression:
\begin{align}
	&\bE_{l_\star}\left[\frac{\csc(\phi_\star) e^{-\frac{\mu}{\pi}\arcsin(\sqrt{1-\cos(\zeta(z))\csc^2(\phi_\star)})}}{\sqrt{1-\cos(\zeta(z)\csc(\phi_\star))}}\right] \nnb\\
	&=\frac{\lambda}{2} \int_{\pi/2-\zeta(z)}^{\pi/2+\zeta(z)}\frac{ e^{-\frac{\mu}{\pi}\arcsin(\sqrt{1-\cos^2(\zeta(z))\csc^2(\phi_\star)})}}{\sqrt{1-\cos^2(\zeta(z)\csc^2(\phi_\star))}}\diff \phi_\star\nnb\\
	&=\lambda \int_{0}^{\zeta(z)} \!\frac{ e^{-\frac{\mu}{\pi}\arcsin\left(\sqrt{1-\cos^2(\zeta(z))\sec^2(v)}\right)}}{\sqrt{1-\cos^2(\zeta(z))\sec^2(v)}}\diff v, \label{48}
\end{align}
where we use the Campbell mean value theorem \cite{baccelli2010stochastic} to integrate the function over the set $(0,\pi)\times (\pi/2-\zeta(z),\pi/2+\zeta(z))$ and the change of variables. 

Then, we integrate w.r.t. $\Xi$ to arrive at 
\begin{align}
	&\bE_\Xi\left[\prod_{\psi_i\in\Xi}^{|\pi/2-\phi_i|<\zeta(z)}e^{-\frac{\mu}{\pi}\arcsin(\sqrt{1-\cos^2(\zeta(z))\csc^2(\phi_i)})}\right]\nnb\\
	&=e^{-\frac{\lambda}{2} \int_{\pi/2-\zeta(z)}^{\pi/2+\zeta(z)} \sin(\phi)\left( 1 - e^{-\frac{\mu}{\pi}\arcsin(\sqrt{1-\cos^2(\zeta(z))\csc^2(\phi)})} \right)\diff \phi }\nnb\\
	&=e^{-{\lambda} \int_{0}^{\zeta(z)} \cos(v)\left( 1 - e^{-\frac{\mu}{\pi}\arcsin(\sqrt{1-\cos^2(\zeta(z))\sec^2(v)})} \right)\diff  v } \label{49}.
\end{align}
	
	Finally, we combine Eqs. \eqref{48}, \eqref{49}, and the rest of variables in Eq. \eqref{eq:45} to obtain the SNR coverage probability of the typical aerial platform.

On the other hand, since the aerial platforms are at the altitude of $h_a$ from the ground nodes, the SNR coverage probability of the typical terrestrial gateway is given by 
\begin{align}
	\bP(\SNR_{U})&=\bP\left(\frac{p_ag_ag_gHh_a^{-\alpha}}{N_oB_a}\geq \tau\right)\nnb\\
	&=\bP\left(H\geq \frac{\tau N_oB_a h_a^\alpha}{p_a g_a g_g}\right)\nnb\\
	&={\overbar{F}_H}\left(\frac{\tau N_oB_a h_a^\alpha}{p_a g_a g_g}\right)={\overbar{F}_H}\left(\frac{\tau h_a^\alpha}{\eta_a}\right), \nnb
\end{align}
where $\eta_a=(p_ag_ag_g/N_0B_a).$

\section{Proof of Theorem \ref{Theorem5}}\label{A:5}
	For the end-to-end communications, the capacity of the proposed network architecture is given by the minimum of the achievable rates of the satellite-to-platform communications and of the platform-to-ground communications. 
\par The achievable rate of the satellite-to-aerial platform communications is  $\bE[\log_2(1+\SNR_{A})]$ where $\SNR_{A}$ is the random variable for the SNR of the typical aerial platform. Since $\log_2(1+\SNR_A)$ is a positive random variable, the expectation of above is given by 
\begin{align}
	\bE[\log_2(1+\SNR_{A})]
	&=\int_0^\infty\bP(\SNR_{A}>  2^u-1)\diff u \label{44},
\end{align}
where the CCDF of the SNR of the typical aerial platform is Eq. \eqref{eq:theorem4} with the variable $\tau$ being replaced with $2^u-1.$ Similarly, the achievable rate of the aerial platform-to-terrestrial gateway is 
\begin{align}
	\bE[\log_2(1+\SNR_{G})]
	&=\int_0^\infty\bP(\SNR_{G}>  2^u-1)\diff u, \label{45}
\end{align}
where the CCDF of SNR  of the typical terrestrial gateway is given by Eq. \eqref{eq:theorem4-1} with the variable $\tau$ being replaced with $2^u-1.$ Finally, the end-to-end capacity is determined by the minimum of the above achievable rates, provided in Eqs. \eqref{44} and \eqref{45}, respectively. 

\section{Proof of Theorem \ref{Theorem:6}}\label{A:6}

		First, the delay is equal to zero if there is any satellite in  $\bS_{A,|\overbar{AC}|}.$ We have 
\begin{align}
	&\bP(T=0) \nnb \\
	&= 1-\bP(\text{no satellite on }\bS_{A,|\overbar{AC}|})\nnb\\
	&=1-e^{\left(-{\lambda}\int_{0}^{\overbar{\varphi}}\cos(\varphi)\left(1-e^{-\frac{\mu }{\pi}\left(\arcsin\left(\sqrt{1-\cos^2(\overbar{\varphi})\csc^2(\phi)}\right)\right)}\right) \right)\diff \phi}\nnb,
\end{align}
where we use the connectivity probability given by Eq. \eqref{eq:theorem2}. 

On the other hand, the typical delay is greater than $t$ if there has been no satellite until the time $t. $ Since the satellites are modeled by the Cox point process conditionally on the orbit process $\Xi$, we have 
\begin{align}
	&\bP(T>{t}) \nnb\\
	&= \bP(\text{no satellite on }\bS_{A,|\overbar{AC}|} \text{ over time } [0,{t}))\nnb\\
	&=\bP\left(\prod_{\Xi}\bP(l(\theta_i,\phi_{i}) \cap \bS_{A,|\overbar{AC}|} =\emptyset \text{ over time $[0,t)$})\right),\nnb
\end{align}
where we use the fact that if there is no satellites on each orbit $l(\theta_i,\phi_i)$ for the duration of time $t,$ each orbit $l(\theta_i,\varphi_i)$ contains no point on the extended spherical cap $\bS_{A,|\overbar{AC}|}$ over the time duration $t$. 

\par Since the angular speed of satellites is $\nu$ and it travels the total length of $r_s\nu t$ over the time $t$, the probability of the above expression corresponds to the probability that the arc of the length $ r_s \nu t + 2 r_s \arcsin(\sqrt{1-\cos^2(\overbar{\varphi}) \csc^2(\phi_i)}) $  has no point. As a result, the delay distribution is given by 
\begin{align}
	&\bP(T>t) \nnb\\&= \bP\left( \prod_{\Xi}^{|\phi_i-\pi/2|<\overbar{\varphi}} e^{-\frac{\mu }{2\pi}\left(\nu t +2\arcsin\left(\sqrt{1-\cos^2(\overbar{\varphi})\csc^2(\phi)}\right)\right)}\right)\nnb\\
	&=e^{ -{\lambda}\int_{0}^{\overbar{\varphi}}\cos(\varphi)\left(1-e^{-\frac{\mu }{2\pi}\left(\nu t +2\arcsin\left(\sqrt{1-\cos^2(\overbar{\varphi})\csc^2(\phi)}\right)\right)}\right) \diff \varphi }, \nnb
\end{align}
where we use the fact that only the orbits of the inclination angles $\phi\in [\pi/2-\overbar{\varphi},\pi/2+\overbar{\varphi}]$ intersect $\bS_{A,|\overbar{AC}|}.$ Then, we use the probability generating functional of the Poisson orbit process and the change of variable to obtain the final result. 
\section{Proof of Proposition \ref{Proposition:1}}\label{A:7}
The typical terrestrial gateway  is located at $(0,0,r_e).$ Conditionally on $Z$---the random elevation angle of the typical aerial platform, the distance from the origin to the platform is $r_l=\sqrt{(r_a)^2+(r_a-r_e)^2\tan^2(z)}$
where $Z=z$ and $r_l\geq r_a$. Then, the visible area from this platform is a spherical cap whose central angle---angle measured from the $z$-axis to the rim of the spherical cap---is given by $\hat{\varphi}=\arccos(r_e/r_l)+\arccos(r_e/r_s)$. Therefore, by using the techniques in Appendix \ref{A:2}, the connectivity probability of the network with an arbitrary elevation angle $Z$ is given by 
\begin{equation}
	1-\bE\left[e^{-\lambda\int_{0}^{\hat{\varphi}}{\cos(\varphi)}\left(1-e^{-\frac{\mu}{\pi}\arcsin\left(\sqrt{1-\cos^2(\overbar{\varphi})\sec^2(\varphi)}\right)}\right)\diff \varphi}\right]\nnb,
	\end{equation}
	where the outer expectation is w.r.t. the elevation angle $Z.$ Using the PDF of $Z$, we obtain the final result.
\section{Proof of Proposition \ref{Proposition:2}}\label{A:8}
Let $y$ be the maximum distance from typical terrestrial gateway  to the visible orbit, based on $\kappa$ the minimum elevation angle of the satellite.  Then, the complementary inclination angle of the orbits visible to the typical terrestrial gateway  is 
\begin{equation}
	\arccos\left({(r_e^2+r_s^2-y^2)}/{(2 r_e r_s)}\right)\label{eq:455}.
\end{equation}
Moreover, based on the simple geometry, the variable $y$ satisfies the following second order expression:
\begin{equation}
	y^2+2r_e\sin(\kappa) y + r_e^2-r_s^2=0.
\end{equation}
From the expression, we obtain a solution: $y=r_e\sin(\kappa)+\sqrt{r_e^2\cos^2(\kappa)+r_s^2}. $ Using it, we obtain the complementary inclination angle of the visible orbit of Eq. \eqref{eq:455}. Then, we have the final result by using the steps in Appendix \ref{A:2}.

\section*{Acknowledgment}
This work is supported by the National Research Foundation of Korea (NRF) Grant funded by the Korean Government through the Ministry of Science and ICT under Grant RS-2024-00334240 and RS-2024-00247682. 

\bibliographystyle{IEEEtran}
\bibliography{ref}

\begin{IEEEbiographynophoto}{Chang-Sik Choi (Member,IEEE)}
Chang-Sik Choi received his B.S. degree (summa cum laude) in Electrical Engineering from Seoul National University and his M.S. and Ph.D. degrees in Electrical Engineering from the University of Texas at Austin. He was a recipient of the Kwanjeong International Scholarship (2012--2017) and the National Science and Technology Scholarship. He has held internships at Apple Inc., Samsung Research America, and M87, and previously worked as a Senior Systems Engineer at Qualcomm Wireless R\&D. He is currently an Assistant Professor at Hongik University in Seoul, South Korea. His research interests include cellular networks, vehicular networks, satellite networks, and communications for autonomous driving.
\end{IEEEbiographynophoto}


\end{document}